\documentclass[12pt,letter,prd,showpacs,nofootinbib,preprintnumbers,superscriptaddress]{revtex4-1}

\usepackage{hyperref}
\newcommand{\beq}{\begin{equation}}
\newcommand{\eeq}{\end{equation}}
\usepackage[pdftex]{graphicx}
\usepackage{xcolor}
\usepackage{amssymb,amsmath,bm,natbib}
\usepackage{appendix}

\usepackage{units}
\usepackage{subfig}
\usepackage{cleveref}
\usepackage{booktabs}
\usepackage{xspace}
\usepackage{float}
\AtBeginDocument{
\heavyrulewidth=.08em
\lightrulewidth=.05em
\cmidrulewidth=.03em
\belowrulesep=.65ex
\belowbottomsep=0pt
\aboverulesep=.4ex
\abovetopsep=0pt
\cmidrulesep=\doublerulesep
\cmidrulekern=.5em
\defaultaddspace=.5em
}
\usepackage{todonotes}
\presetkeys{todonotes}{line,color=blue!30, size=\tiny}{}
\newcommand{\nexp}{PIONEER}

\usepackage{amsmath}

\usepackage{siunitx}
\usepackage[ampersand]{easylist}
\usepackage{gensymb}
\usepackage[T1]{fontenc}

\newcommand{\be}{\begin{equation}}
\newcommand{\ee}{\end{equation}}
\newcommand{\bl}{\vspace{1mm}\begin{easylist}}
\newcommand{\el}{\end{easylist}\vspace{1mm}}

\newcommand{\pdar}{\mbox{$\pi$DAR}}

\newcommand{\pdif}{\mbox{$\pi$DIF}}
\newcommand{\mdif}{\mbox{$\mu$DIF}}
\newcommand{\pie}{$\pi\to e \nu$}
\newcommand{\pme}{$\pi\to\mu\to e$}
\newcommand{\app}{\mbox{$\approx$}}
\newcommand{\remu}{\mbox{$R_{e/\mu}$}}
\newcommand{\tmu}{\mbox{4.1\,MeV}}
\newcommand{\gbl}{\mbox{G4Beamline}}
\newcommand{\atar}{\mbox{ATAR}}
\newcommand{\calo}{\mbox{CALO}}
\newcommand{\track}{\mbox{TRACKER}}
\newcommand{\beam}{\mbox{BEAM}}

\newcommand{\thcal}{\mbox{58\,MeV}}

\newcommand{\vud}{\ensuremath{\left|V_{ud}\right|}\xspace}
\newcommand{\vus}{\ensuremath{\left|V_{us}\right|}\xspace}

\usepackage[margin=0pt,font=normalsize,labelfont=bf,labelsep=endash,justification=RaggedRight]{caption}
\begin{document}
\presetkeys{todonotes}{disable}
%\listoftodos
\newpage

% Title
%\title{PSI Ring Cyclotron Proposal\\ PIONEER: Studies of Rare Pion Decays}
\title{Testing Lepton Flavor Universality and CKM Unitarity with Rare Pion Decays 
in the PIONEER experiment}

%\author {PIONEER Collaboration}

%\input{Planning/alphCollaboration}

\author{W.~Altmannshofer}\affiliation {Santa Cruz Institute for Particle Physics (SCIPP), University of California Santa Cruz, 1156 High street, Santa Cruz (CA) 95064 USA}
\author{H.~Binney}\affiliation{Department of Physics, University of Washington, Box 351560, Seattle, Washington 98195 USA }
\author{E.~Blucher}\affiliation{Enrico Fermi Institute and Department of Physics, University of Chicago,
5720 South Ellis Avenue,
Chicago, IL 60637 USA}
\author{D.~Bryman}\affiliation{Department of Physics \& Astronomy, University of British Columbia 6224 Agricultural Road, Vancouver V6T 1Z1 Canada}\affiliation{TRIUMF, 4004 Wesbrook Mall, Vancouver V6T 2A3 Canada}
\author{L.~Caminada}\affiliation{Paul Scherrer Institute, 5232 Villigen PSI Switzerland}
\author{S.~Chen}\affiliation { Department of Engineering Physics, Tsinghua University, 30 Shuangqing Road, Haidian District, Beijing, 100084 P. R. China   }
\author{V.~Cirigliano}\affiliation{Institute for Nuclear Theory, University of Washington, Seattle WA 98195-1550 USA}
\author{S.~Corrodi}\affiliation{Argonne National Laboratory, High Energy Physics Division, 9700 S Cass Ave, Lemont, IL 60439 USA}
\author{A. Crivellin}\affiliation{Paul Scherrer Institute, 5232 Villigen PSI Switzerland}\affiliation{Physik-Institut
University of Zurich 
Winterthurerstrasse  190
CH-8057 Zurich
Switzerland}\affiliation{Division of Theoretical Physics, CERN,Espl. des Particules 1, 1211 Meyrin  Switzerland}
\author{S.~Cuen-Rochin}\affiliation{Tecnol\'{o}gico de Monterrey, School of Engineering and Sciences, Blvd. Pedro Infante 3773 Pte, Culiacan 80100 Mexico}
\author{A.~Di~Canto}\affiliation{Physics Department, Brookhaven National Laboratory, Upton, NY, 11973 USA}
\author{L.~Doria}\affiliation{PRISMA$^+$ Cluster of Excellence and Johannes Gutenberg Universit\"at Mainz, Institut für Kernphysik, J.-J.-Becher-Weg 45, 55128 Mainz Germany}
\author{A.~Gaponenko}\affiliation{Fermi National Accelerator Laboratory (FNAL), P.O. Box 500, Batavia IL 60510-5011 USA}
\author{A.~Garcia}\affiliation{Department of Physics, University of Washington, Box 351560, Seattle, Washington 98195 USA }
\author{L.~Gibbons}\affiliation{Department of Physics, Cornell University, 109 Clark Hall, Ithaca, New York 14853 USA}
\author{C.~Glaser}\affiliation{Department of Physics,  University of Virginia,  P.O. Box 400714, 382 McCormick Road, Charlottesville, VA 22904-4714 USA}
\author{M.~Escobar~Godoy}\affiliation {Santa Cruz Institute for Particle Physics (SCIPP), University of California Santa Cruz, 1156 High street, Santa Cruz (CA) 95064 USA}
\author{D.~G\"oldi}\affiliation{ETH Zurich, Main building, Rämistrasse 101, 8092 Zurich Switzerland}
\author{S.~Gori}\affiliation {Santa Cruz Institute for Particle Physics (SCIPP), University of California Santa Cruz, 1156 High street, Santa Cruz (CA) 95064 USA}
\author{T.~Gorringe}\affiliation{Department of Physics and Astronomy, University of Kentucky, 505 Rose Street Lexington, Kentucky 40506-0055 USA}
\author{D.~Hertzog}\affiliation{Department of Physics, University of Washington, Box 351560, Seattle, Washington 98195 USA }
\author{Z.~Hodge}\affiliation{Department of Physics, University of Washington, Box 351560, Seattle, Washington 98195 USA }
\author{M.~Hoferichter}\affiliation{Albert Einstein Center for Fundamental Physics, Institute for Theoretical Physics, University of Bern, Sidlerstrasse 5, 3012 Bern  Switzerland}
\author{S.~Ito}\affiliation{KEK, High Energy Accelerator Research Organization, 1-1, Oho, Tsukuba-city, Ibaraki 305-0801 Japan}
\author{T.~Iwamoto}\affiliation{International Center for Elementary Particle Physics (ICEPP), The University of Tokyo, 7-3-1 Hongo, Bunkyo-ku, Tokyo 113-0033 Japan}
\author{P.~Kammel}\affiliation{Department of Physics, University of Washington, Box 351560, Seattle, Washington 98195 USA }
\author{B.~Kiburg}\affiliation{Fermi National Accelerator Laboratory (FNAL), P.O. Box 500, Batavia IL 60510-5011 USA}
\author{K.~Labe}\affiliation{Department of Physics, Cornell University, 109 Clark Hall, Ithaca, New York 14853 USA}
\author{J.~LaBounty}\affiliation{Department of Physics, University of Washington, Box 351560, Seattle, Washington 98195  USA }
\author{U.~Langenegger}\affiliation{Paul Scherrer Institute, 5232 Villigen PSI Switzerland}
\author {C.~Malbrunot}\affiliation{TRIUMF, 4004 Wesbrook Mall, Vancouver V6T 2A3 Canada}
\author{S.M.~Mazza}\affiliation {Santa Cruz Institute for Particle Physics (SCIPP), University of California Santa Cruz, 1156 High street, Santa Cruz (CA) 95064 USA}
\author{S.~Mihara}\affiliation{KEK, High Energy Accelerator Research Organization, 1-1, Oho, Tsukuba-city, Ibaraki 305-0801 Japan}
\author{R.~Mischke}\affiliation{TRIUMF, 4004 Wesbrook Mall, Vancouver V6T 2A3 Canada}
\author{T.~Mori}\affiliation{International Center for Elementary Particle Physics (ICEPP), The University of Tokyo, 7-3-1 Hongo, Bunkyo-ku, Tokyo 113-0033 Japan}
\author{J.~Mott}\affiliation{Fermi National Accelerator Laboratory (FNAL), P.O. Box 500, Batavia IL 60510-5011 USA}
\author{T.~Numao}\affiliation{TRIUMF, 4004 Wesbrook Mall, Vancouver V6T 2A3 Canada}
\author{W.~Ootani}\affiliation{International Center for Elementary Particle Physics (ICEPP), The University of Tokyo, 7-3-1 Hongo, Bunkyo-ku, Tokyo 113-0033 Japan}
\author{J.~Ott}\affiliation {Santa Cruz Institute for Particle Physics (SCIPP), University of California Santa Cruz, 1156 High street, Santa Cruz (CA) 95064 USA}
\author{K.~Pachal}\affiliation{TRIUMF, 4004 Wesbrook Mall, Vancouver V6T 2A3 Canada}
\author{C.~Polly}\affiliation{Fermi National Accelerator Laboratory (FNAL), P.O. Box 500, Batavia IL 60510-5011 USA}
\author{D.~Po\v{c}ani\'c}\affiliation{Department of Physics, University of Virginia, P.O. Box 400714, 382 McCormick Road, Charlottesville, VA 22904-4714 USA}
\author{X.~Qian}\affiliation{Physics Department, Brookhaven National Laboratory, Upton, NY, 11973 USA}
\author{D.~Ries}\affiliation{Department of Chemistry – TRIGA site, Johannes Gutenberg University Mainz, Fritz-Strassmann-Weg 2, 55128 Mainz Germany}
\author{R.~Roehnelt}\affiliation{Department of Physics, University of Washington, Box 351560, Seattle, Washington 98195 USA }
\author{B.~Schumm}\affiliation {Santa Cruz Institute for Particle Physics (SCIPP), University of California Santa Cruz, 1156 High street, Santa Cruz (CA) 95064 USA}
\author{P.~Schwendimann}\affiliation{Department of Physics, University of Washington, Box 351560, Seattle, Washington 98195 USA }
\author{A.~Seiden}\affiliation {Santa Cruz Institute for Particle Physics (SCIPP), University of California Santa Cruz, 1156 High street, Santa Cruz (CA) 95064 USA}
\author{A.~Sher}\affiliation{TRIUMF, 4004 Wesbrook Mall, Vancouver V6T 2A3 Canada}
\author{R.~Shrock}\affiliation {C. N. Yang Institute for Theoretical Physics and Department of Physics and Astronomy, Stony Brook University, Stony Brook, NY 11794 USA}
\author{A.~Soter}\affiliation{ETH Zurich, Main building, Rämistrasse 101, 8092 Zurich Switzerland}
\author{T.~Sullivan}\affiliation{Department of Physics and Astronomy, Elliott Building, University of Victoria, Victoria, BC V8P 5C2 Canada}
\author{M.~Tarka}\affiliation {Santa Cruz Institute for Particle Physics (SCIPP), University of California Santa Cruz, 1156 High street, Santa Cruz (CA) 95064 USA}
\author{V.~Tischenko}\affiliation{Physics Department, Brookhaven National Laboratory, Upton, NY, 11973 USA}
\author{A.~Tricoli}\affiliation{Physics Department, Brookhaven National Laboratory, Upton, NY, 11973 USA}
\author{B.~Velghe}\affiliation{TRIUMF, 4004 Wesbrook Mall, Vancouver V6T 2A3 Canada}

\author{V.~Wong}\affiliation{TRIUMF, 4004 Wesbrook Mall, Vancouver V6T 2A3 Canada}
\author{E.~Worcester}\affiliation{Physics Department, Brookhaven National Laboratory, Upton, NY, 11973 USA}
\author{M.~Worcester}\affiliation{Instrumentation Division, Brookhaven National Laboratory, Upton, NY, 11973 USA}
\author{C.~Zhang}\affiliation{Physics Department, Brookhaven National Laboratory, Upton, NY, 11973 USA}

%Abstract
\begin{abstract}
ABSTRACT: The physics motivation and the conceptual design of the PIONEER experiment, a next-generation 
rare pion decay experiment testing lepton flavor universality and CKM unitarity, are described. 
Phase I of the PIONEER experiment, which was proposed and approved at Paul Scherrer Institut,
aims at measuring the charged-pion branching ratio to electrons vs.\ muons, $R_{e/\mu}$, 
15 times more precisely than the current experimental result, reaching the precision of the Standard Model 
(SM) prediction at 1 part in $10^4$. Considering several inconsistencies between the SM predictions 
and data pointing towards the potential violation of lepton flavor universality, the PIONEER experiment will 
probe non-SM explanations of these anomalies through sensitivity to quantum effects of new particles
up to the PeV mass scale. 
The later phases of the PIONEER experiment aim at improving the experimental precision of the branching ratio 
of  pion beta decay (BRPB), $\pi^+\to \pi^0 e^+ \nu (\gamma)$, currently at 
$1.036(6)\times10^{-8}$, by a factor of three (Phase II) and 
an order of magnitude (Phase III).  Such precise measurements of BRPB will allow for tests of CKM unitarity in 
light of the Cabibbo Angle Anomaly and the theoretically cleanest extraction of $|V_{ud}|$ at 
the 0.02\% level, comparable to the deduction from superallowed beta decays. 
\end{abstract}
%\date{\today}
\maketitle
%\newpage

  \begin{center} 
\rule[-0.2in]{\hsize}{0.01in}\\
  \rule{\hsize}{0.01in}\\
  \vskip 0.01in
  Submitted to the Proceedings of the US Community Study \\
on the Future of Particle Physics (Snowmass 2021)\\
  \rule{\hsize}{0.01in}\\
  \rule[+0.2in]{\hsize}{0.01in}\\[-2em]
\end{center}
%{\bf Beam and area requirements:}
%\begin{itemize}
%\item    Beam line $\pi E5$; QSK quadrupole triplet.
%\item Area: Downstream of the SB43 quadrupole magnet between
%the concrete wall and the current MEG II infrastructure.
%\item Low momentum separator and collimater; tentatively SEP41 $E \times B$ Wien
%Filter for particle separation.  
%\item Electrical power: $\leq 200$kW.
%\item   Beam properties: $\pi^+$; $50-75$ MeV/c with $\Delta p/p=2\%$; Phase I rate: $3\times10^5$ Hz;
%$\mu, e$ contamination $<10\%$.
%\item    Duration of the experiment: 3 years.
%\item Beam time request for the first beam period after approval: see separate document.
%    \end{itemize}
%\newpage
%{\bf Hazardous Material/Equipment}
%\begin{itemize}
%\item Liquid xenon calorimeter; 3000 l.
%    \end{itemize}
%\begin{figure}[htbp]
%\centering
%\includegraphics[width=1.0\textwidth]{Figures/HazzardTable.pdf}
%\caption{ }
%\label{}
%\end{figure}
%\newpage
\vspace{-1cm}
\tableofcontents
%\addtocontents{toc}{\vskip-40pt}

\newpage
\section{Executive Summary}
In recent years, there have been an increasing number of intriguing hints for lepton flavor universality violation
(LFUV). Motivated by these indications of physics beyond the Standard Model, Phase I of the PIONEER experiment, approved at the Paul Scherrer Institute (PSI),
aims to measure the charged-pion branching ratio to electrons vs.\ muons $R_{e/\mu}$ to 1 part in $10^4$, 
improving the current experimental result $ R_{e/\mu} \hspace{0.1cm}\text{(exp)} =1.2327(23)\times10^{-4}$~\cite{ParticleDataGroup:2020ssz,PiENu:2015seu,Bryman:1985bv,Britton:1993cj,Czapek:1993kc}
by a factor of 15. This precision on $R_{e/\mu}$ will match the theoretical accuracy of the SM prediction allowing 
for a test of LFU at an unprecedented level, 
probing non-SM explanations of existing LFUV anomalies 
through sensitivity to quantum effects of new particles up to the PeV mass scale.

Phase II and III of PIONEER experiment aim to improve the experimental precision of the branching ratio of pion 
beta decay, $\pi^+\to \pi^0 e^+ \nu (\gamma)$, currently at $1.036(6)\times10^{-8}$, by a factor 
of three and an order of magnitude, respectively. The improved measurements of pion beta decay would allow 
one to extract \vud in a theoretically pristine manner.
%, which is a critical input in evaluating the 
%significance of CAA. 
%The precision of \vud will be improved by a factor of 3 in Phase II when combined with $K_{\ell 3}$ 
%decays~\cite{Czarnecki:2019iwz} and an order of magnitude for a stand-alone extraction in Phase III. 
The ultimate precision of \vud is expected to reach the  0.02\% level, which is comparable to the currently 
most precise deduction from superallowed beta decays, allowing for a stringent test of CKM unitarity. Furthermore, the PIONEER experiment will also improve the experimental 
limits~\cite{Bryman:2019ssi,Bryman:2019bjg,PIENU:2021clt} by an order of magnitude or more to a host of exotic decays probing for effects of 
heavy neutrinos and dark sector physics.

%\begin{figure}[H]
%\centering
% \includegraphics[scale=.5]{Figures/GenericDetector.png}
%\includegraphics[width=0.4\textwidth]{Figures/Simulation/pioneer_drawing_largerfonts.pdf}
%\caption{Layout of the PIONEER rare pion decay experiment.  The intense positive pion beam enters from the left and   is brought to rest in a highly segmented active target (ATAR).   Decay positron trajectories are measured from the ATAR to an outer electromagnetic calorimeter (CALO) through a tracker.  The CALO records the positron energy, time and location.
%with  high resolution, and a hit position with modest resolution.
%}
%\label{fig:GenericDetector}
%\end{figure}

%ETW comment: I think we should describe the measurement strategy here, but not the details of the detector. Will put back some of this
%As shown in Fig.~\ref{fig:GenericDetector}, 
The conceptual design of the Phase-I PIONEER experiment 
includes a 3$\pi$-sr 25 radiation length liquid xenon calorimeter, a segmented low gain avalanche detector (LGAD) stopping target, a positron tracker, 
and other detectors. Compared to the previous generation of rare pion decay experiments, the 4-D (position and time) tracking capability of the LGAD-based active target allows 
for excellent separation of $\pi \rightarrow  e \nu$ signal from vast amount of $\pi\to\mu\to e$ background ($\pi\to \mu\nu$ followed by $\mu \rightarrow e \nu  \overline{\nu}$).
%$\pi \rightarrow \nu  \left(\mu \rightarrow e \nu  \overline{\nu} \right)$ .

%using pattern recognition
%techniques in addition to the energy and timing information in an intense pion beams. At the same time, the 
%fast scintillation light response and the large acceptance of LXe calorimeter enables a high statistics 
%measurement of $\pi \rightarrow e + \nu$ decays, while the deep and uniform LXe calorimeter form a solid
%foundation for excellent energy resolution. The excellent time and energy resolutions, greatly increased calorimeter depth, high-speed detector and electronics response, large solid angle coverage, and complete event reconstruction  are all critical aspects of the design of PIONEER experiment.

The PIONEER collaboration consists of participants from both the nuclear and particle physics communities
including PIENU, PEN/PiBeta, and MEG/MEGII collaborations, as well as experts in rare kaon decays, 
low-energy stopped muon experiments, the muon $g-2$ experimental campaign, high energy collider physics, 
neutrino physics, and other areas. The collaboration plans R\&D in several critical areas including i) beam studies, ii) LGAD-based active target (sensor and readout 
electronics), iii) LXe calorimeter (optical sensor and optical segmentation), iv) DAQ, and v) trigger, in preparation 
for a technical design report. The collaboration is still developing and welcomes new members.

This Snowmass white paper describes the physics motivation and the conceptual design of the PIONEER
experiment, and is prepared based on the PIONEER proposal~\cite{pioneer_proposal} submitted to
and approved with  high priority by the PSI (CHRISP - Swiss Research InfraStructure for Particle physics) program advisory 
committee. Using intense pion beams, and state-of-the-art instrumentation and computational resources, the PIONEER 
experiment is aiming to take data at PSI by the end of this decade.

\newpage
\section{Introduction}

Precise low-energy measurements of observables that can be very accurately calculated in the Standard Model (SM)
offer highly sensitive tests of new physics (NP). In light of the existing intriguing hints for lepton flavor
universality (LFU) violating NP~\cite{Crivellin:2021sff,Fischer:2021sqw,Bryman:2021teu}, the ratio 
$R_{e/\mu} = \Gamma(\pi^+\rightarrow e^+\nu(\gamma))/\Gamma(\pi^+\rightarrow \mu^+\nu(\gamma))$ for pion 
decays to positrons relative to muons is especially promising: %, known theoretically with a precision better than $\unit[0.01]{\%}$,
it is one of the most precisely known observables involving quarks within the SM and NP can even 
have (chirally) enhanced effects, making it an extremely sensitive probe of NP.
%The branching ratio $R_{e/\mu} = \Gamma(\pi^+\rightarrow \mu+\nu(\gamma))/\Gamma(\pi^+\rightarrow e+\nu(\gamma))$ for pion decays to positrons over muons is a highly sensitive probe for new physics. In the Standard Model (SM), $R_{e/\mu}$ has been calculated with extraordinary precision at the $10^{-4}$ level perhaps the most precisely calculated weak interaction observable involving quarks. 
However, while the uncertainty of the SM calculation for $R_{e/\mu}$ is 
very small (with relative precision $1.2\times 10^{-4}$~\cite{Cirigliano:2007xi}), the current experimental world average is about a factor 15 less precise, limiting the NP reach.
%and the decay $\pi^+\rightarrow e^+\nu$ is helicity suppressed by the $V-A$ structure of charged currents, a measurement of $R_{e/\mu}$ is extremely sensitive to the presence of pseudoscalar (and scalar) couplings absent from the SM; a disagreement with the theoretical expectation would unambiguously imply the existence of NP. Moreover, the comparison of $R_{e/\mu}$ measurements with SM theory already now provides the most accurate measure of lepton universality at the level of $0.1\%$.

A new experiment, PIONEER, is proposed at the Paul Scherrer Institute (PSI), where the highest intensity low energy pion beams 
are delivered. The Phase~I of PIONEER~\cite{pioneer_proposal}, which has been approved by PSI (CHRISP - Swiss Research InfraStructure for Particle physics) program advisory committee with  high priority, will bridge the gap of a factor 15  between theoretical and experimental precision for $R_{e/\mu}$. With measurements at the $0.01\%$ level in precision, NP up to the PeV scale~\cite{Bryman:2011zz} may be revealed. Such precision would  contribute to  stringent tests of LFU in a context where several intriguing hints of LFU violation (LFUV) have emerged. In addition, it will allow extended searches for exotics such as heavy neutral leptons and dark sector processes. In later phases (II, III),  PIONEER  will  also study  pion beta decay $\pi^+\to \pi^0 e^+ \nu (\gamma)$ ultimately aiming at  an order of magnitude improvement in precision to determine \vud in a theoretically pristine manner and test CKM unitarity, for which there is presently a $\approx3\sigma$ tension~\cite{ParticleDataGroup:2020ssz}. PIONEER is an ambitious program that will span more than a decade of research activity at PSI. 
%With the current proposal we seek approval of the Phase I measurement of $R_{e/\mu}$, so that we can
%move forward with applications to  national funding agencies.
While we focus on the measurement of the \pie~ branching ratio $R_{e/\mu}$ in this paper, the following 
sections discuss the theoretical motivation for pursuing the full rare pion decay program.  
Discussions of the \nexp\ detector concepts, simulations, estimated sensitivities, and planning for realization follow. 
%In the final section, we discuss aspects related to training, equity, diversity and inclusion.
%Appendices contain greater detail on some of the topics.

\section {Motivation}
%\subsection{Theory Overview}

\begin{figure}[t]
\centering
%\begin{subfigure}[t!]{0.4\textwidth}
%\hspace{-0.5cm}
\includegraphics[width=0.9\textwidth]{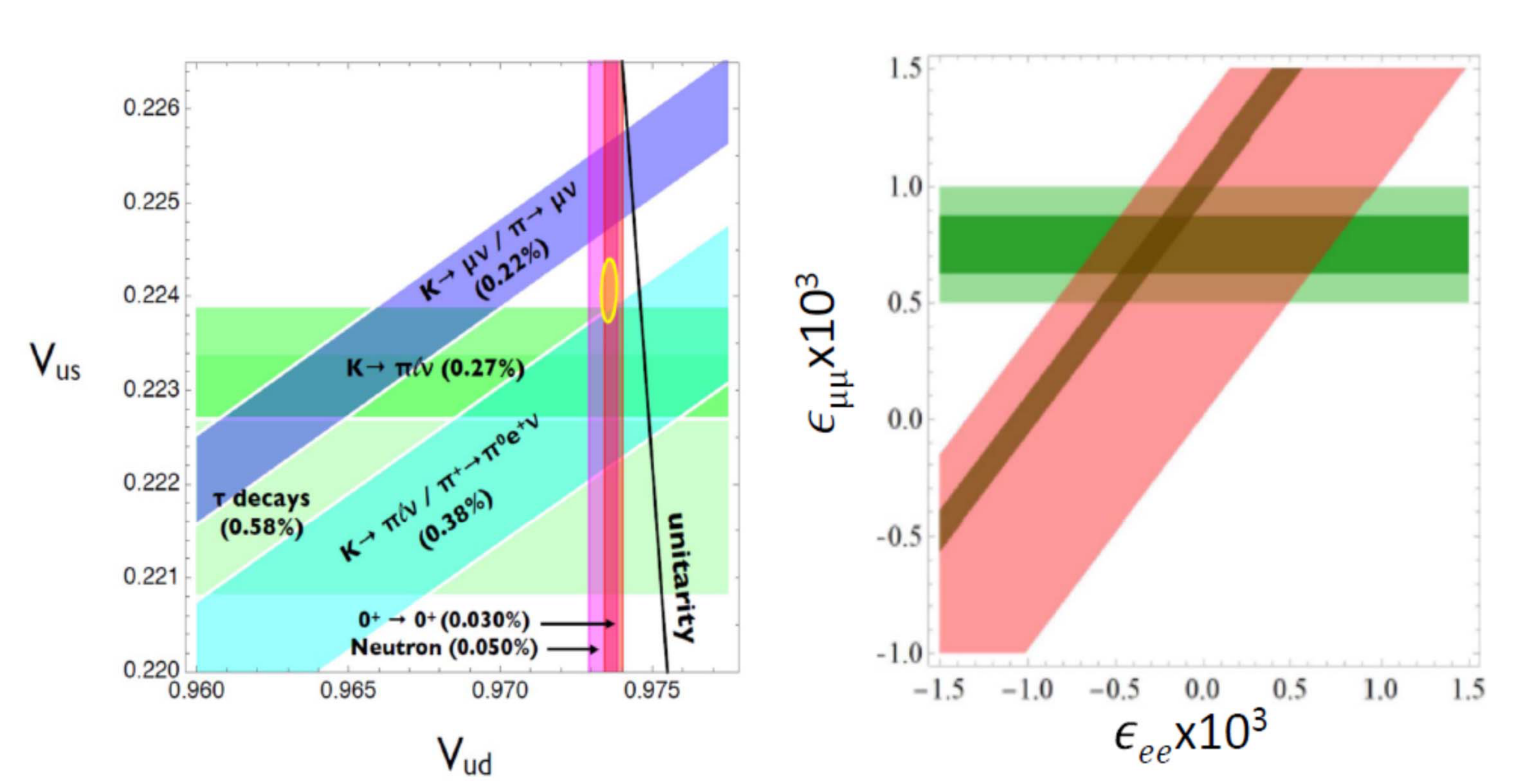}
%
%\includegraphics[width=0.6\textwidth]{Figures/fig-ckm.pdf}
%
%\caption{Existing tensions in the 1st-row CKM unitarity test.}
%\end{subfigure}
%\begin{subfigure}[t!]{0.4\textwidth}
%\includegraphics[scale=0.5]{sections/figures/Fit.pdf}
%
%\hspace{-9mm}
%\includegraphics[width=0.4\textwidth]{Figures/Fit.jpg}
%
\caption{Left: Tensions in the first-row CKM unitarity test (see text)~\cite{Bryman:2021teu}. Right: Constraints ($1\sigma$) on modified $W\ell\nu$ couplings from CKM unitarity (green) and LFUV (red) (adapted from Ref.~\cite{Crivellin:2020lzu}). The light bands  show the current status and the dark bands include the expected PIONEER  sensitivity. The SM values of $A_\ell$ are assumed to be modified by $1+\epsilon_{\ell \ell}$ where $\ell=e,\mu$.}
\label{fig:CKM}
%\label{fig:Fit_epsii}
%\end{subfigure}
\end{figure}

While no particles or interactions beyond those of the SM have been observed so far, intriguing hints for LFUV have been accumulated in recent years~\cite{Crivellin:2021sff,Fischer:2021sqw,Bryman:2021teu}. In particular, the measurements of the ratios of branching ratios (Br)  $R(D^{(*)})=Br[ B\to D^{(*)}\tau\nu_\tau$]/Br[$B\to D^{(*)}\ell\nu_\ell$]~\cite{Lees:2012xj,Aaij:2017deq,Abdesselam:2019dgh} , where $\ell=\mu, e$, and  $R(K^{(*)})=Br[B\to K^{(*)} \mu^+ \mu^-$]$/$Br[$B\to K^{(*)} e^+ e^-$]~\cite{Aaij:2017vbb,LHCb:2019hip,LHCb:2021trn} deviate from the SM expectation by more than $3\sigma$~\cite{Amhis:2019ckw,Murgui:2019czp,Shi:2019gxi,Blanke:2019qrx,Kumbhakar:2019avh} and $4\sigma$~\cite{Alguero:2019ptt,Aebischer:2019mlg,Ciuchini:2019usw,Arbey:2019duh}, respectively. In addition, anomalous magnetic moments $(g-2)_\ell$ ($\ell=e,\mu,\tau$) of charged leptons are intrinsically related to LFUV, as they are chirality flipping quantities. Here, the longstanding discrepancy in $(g-2)_\mu$, just reaffirmed at the level of $4.2\sigma$~\cite{Bennett:2006fi,Muong-2:2021ojo,Aoyama:2020ynm}, can be considered as another hint of LFUV, since, if compared to $(g-2)_e$, the NP contribution scales with a power of the lepton mass~\cite{Davoudiasl:2018fbb,Crivellin:2018qmi}.  In addition, there is a hint for LFUV in the difference of the forward-backward asymmetries ($\Delta A_{\rm FB}$) in $B\to D^*\mu\nu$ vs $B\to D^*e\nu$~\cite{Bobeth:2021lya,Carvunis:2021dss}. As another possible indication of LFUV, CMS observed an excess in non-resonant di-electron pairs with respect to di-muons~\cite{Sirunyan:2021khd}. Furthermore, the possible deficit in first-row unitarity of the Cabibbo-Kobayashi-Maskawa (CKM) matrix, known as the Cabibbo angle anomaly (CAA) (see Fig.~\ref{fig:CKM} (left)), can also be viewed as a sign of LFUV~\cite{Coutinho:2019aiy,Crivellin:2020lzu}. For these reasons, there is very strong motivation for an upgraded $R_{e/\mu}$ experiment whose precision matches that of the SM prediction.

The branching ratio 
%\begin{equation}
$R_{e/\mu} = \frac{\Gamma\left(\pi^+ \rightarrow e^+ \nu (\gamma) \right)}{\Gamma\left(\pi^+ \rightarrow \mu^+ \nu (\gamma)\right)}$
%\end{equation}
for pion decays to electrons over muons provides the best test of electron--muon universality in charged-current weak interactions. In the SM, $R_{e/\mu}$ has been calculated with extraordinary precision at the $10^{-4}$ level as \cite{Cirigliano:2007xi,Cirigliano:2007ga,Marciano:1993sh}
\begin{equation}
\label{Remu_SM}
    R_{e/\mu} \hspace{0.1cm}\text{(SM)} = 1.23524(15)\times10^{-4},
\end{equation}
perhaps the most precisely calculated weak interaction observable involving quarks.\footnote{Reference~\cite{Bryman:2011zz} estimates the uncertainty due to the unknown non-leading-logarithmic contributions of $O(\alpha^2 \log (m_\mu/m_e)$ in a different way compared to Ref.~\cite{Cirigliano:2007xi}. This leads to a larger total uncertainty, i.e.,  $R_{e/\mu} \hspace{0.1cm}\text{(SM)} = 1.23524(19)\times10^{-4}$.}
In comparison, the current experimental precision~\cite{ParticleDataGroup:2020ssz,PiENu:2015seu,Bryman:1985bv,Britton:1993cj,Czapek:1993kc}
\begin{equation}
   R_{e/\mu} \hspace{0.1cm}\text{(exp)} =1.2327(23)\times10^{-4}.
\end{equation}
is more than a factor of 15 worse. Because the
uncertainty of the SM calculation for $R_{e/\mu}$ is very small and the decay $\pi^+ \rightarrow e^+ \nu$ is helicity-suppressed by the $V-A$ structure of charged currents, a measurement of $R_{e/\mu}$ is extremely sensitive to the presence of pseudoscalar (and scalar) couplings absent from the SM. Comparison between $R_{e/\mu}$ in theory and experiment provides a stringent test of the $e$--$\mu$ universality of the weak interaction; a disagreement with the theoretical expectation would unambiguously imply the existence of NP. With the  PIONEER Phase I physics goal of improving  
$R_{e/\mu}$ experimental precision by a factor of 15 to 0.01\% level, NP at the PeV scale can be probed~\cite{Bryman:2011zz}, even up to several PeV in specific models such as leptoquarks. 
%an order of magnitude beyond the current experimental precision~\cite{ParticleDataGroup:2020ssz,PiENu:2015seu,Bryman:1985bv,Britton:1993cj,Czapek:1993kc}
%\begin{equation}
%   R_{e/\mu} \hspace{0.1cm}\text{(exp)} %=1.2327(23)\times10^{-4}.
%\end{equation}
%$R_{e/\mu}$ thus provides a unique opportunity for a pristine test of LFU in the quark sector. 
%(see Sec.~\ref{sec:LFUV}). 
Assuming that LFUV originates from modified $W\ell\nu$ couplings, the determination of CKM elements
will also be affected. Importantly, beta decays have an enhanced sensitivity to a modified $W\mu\nu$
coupling, due to a CKM enhancement by $\left|V_{ud} / V_{us}\right|^2 \sim 20$. Such a modification
of the $W\ell\nu$ couplings would also affect $R_{e/\mu}$, albeit for a different flavor combination (see Fig.~\ref{fig:CKM} (right)). This connection provides further motivation for an improved $R_{e/\mu}$ measurement, especially because the sensitivity to LFUV would be comparable to future improved constraints from beta decays. 
%Moreover, recent global fits to electroweak observables and tests of LFU show a preference for $R_{e/\mu}$ larger than its SM expectation \cite{Coutinho:2019aiy,Crivellin:2020ebi}. Furthermore, $R(K^{*})$ can be correlated to $R_{e/\mu}$~\cite{Capdevila:2020rrl} and a combined explanation of the deficit in the first-row CKM unitarity and the CMS excess in di-electrons even predicts that $R_{e/\mu}$ should be larger than its SM value~\cite{Crivellin:2021rbf}.

The significance of the CAA, a $3\sigma$ tension with the CKM unitarity illustrated in 
Fig.~\ref{fig:CKM} (left)~\cite{Bryman:2021teu,Cirigliano:2021yto}, depends crucially on 
experimental input quantities used for the extraction of CKM matrix elements as well as a 
number of theory corrections.
The detector optimized for a next-generation $R_{e/\mu}$ experiment will also be ideally suited for a high-precision measurement of pion beta decay, which allows one to extract \vud in a theoretically pristine manner.
The branching ratio for pion beta decay was most accurately measured by the PiBeta experiment at PSI
\cite{Pocanic5,Frlez:2003vg,Pocanic:2003pf,Frlez:2003pe,Bychkov:2008ws} to be
\begin{equation}\frac{\Gamma(\pi^+ \to  \pi^0 e^+ \nu)}{\Gamma(\text{Total})}= 1.036 \pm 0.004 ~(\text{stat}) \pm 0.004~(\text{syst}) \pm 0.003~(\pi\to e\nu) \times 10^{-8},
\end{equation}
where the first uncertainty is statistical, the second systematic, and the third is the $\pi\to e\nu$ branching ratio uncertainty. While the pion beta decay provides the theoretically cleanest determination of the magnitude of the CKM matrix element \vud, the current extraction of
$\vud = 0.9739(28)_{\textrm{exp}}(1)_{\textrm{th}}$ at 0.3\%  is not presently relevant for the CKM unitarity tests because superallowed nuclear beta decays provide a nominal precision of 
0.03\%. In order to make $\pi^+ \rightarrow \pi^0 e^+ \nu (\gamma)$ important for CKM unitarity tests, a two-step improvement in experimantal precision is identified. As advocated in Ref.~\cite{Czarnecki:2019iwz}, the first step is a three-fold improvement in BRPB precision 
compared to Ref.~\cite{Pocanic:2014jka} as proposed in PIONEER Phase II 
would allow for a 0.2\% determination of $\left|V_{us}/V_{ud}\right|$ via 
improving the measurement of the ratio 
  \begin{equation}
   \label{rv}
        R_V = \frac{\Gamma\left(K \rightarrow \pi l \nu (\gamma) \right)}{\Gamma\left(\pi^+ \rightarrow \pi^0 e^+ %%\nu (\gamma)\right)}=1.9884(115)_{\pi}(42)_K\times %10^7,
\nu (\gamma)\right)}=1.9884(115)_{\pi}(93)_K\times 10^7,    
\end{equation} 
when combined with $K_{\ell 3}$ decays. 
As shown in Fig.~\ref{fig:CKM}, this would match the precision of the current extraction of $\left|V_{us} / V_{ud}\right|$ from  the axial channels~\cite{Marciano:2004uf} via 
\begin{equation}
        R_A = \frac{\Gamma\left(K \rightarrow \mu \nu (\gamma) \right)}{\Gamma\left(\pi \rightarrow \mu \nu (\gamma)\right)}=1.3367(25),
\end{equation}
thus providing a new competitive constraint on the \vus--\vud plane and 
probing NP that might affect vector and axial-vector channels in different ways. 
In the second step, an order of magnitude improvement  in the BRPB precision will be sought
in PIONEER Phase III program. This would provide the theoretically cleanest extraction of \vud at the 0.02\% level, comparable to the current value from superallowed beta decays~\cite{Hardy:2020qwl}.

Finally, PIONEER will improve sensitivity to a host of exotic decays, including probes for the effects of heavy neutrinos~\cite{Shrock:1980vy,Shrock:1980ct,Abela:1981nf,Minehart:1981fv,Bryman:1983cja,Azuelos:1986eg,Britton:1992pg, PiENu:2015seu,PIENU:2017wbj,PIENU:2019usb, Bryman:2019ssi,Bryman:2019bjg, PIENU:2021clt}
, unique capabilities to search for pion decays to various light dark sector 
particles~\cite{Altmannshofer:2019yji,Dror:2020fbh,Batell:2017cmf,PIENU:2021clt},
and lepton-flavor-violating decays of the muon into light NP 
particles $\mu \to e X$.

\section{PIONEER Experiment}
\subsection{Experiment Overview and Strategy}

\begin{figure}[H]
\centering
\includegraphics[width=0.4\textwidth]{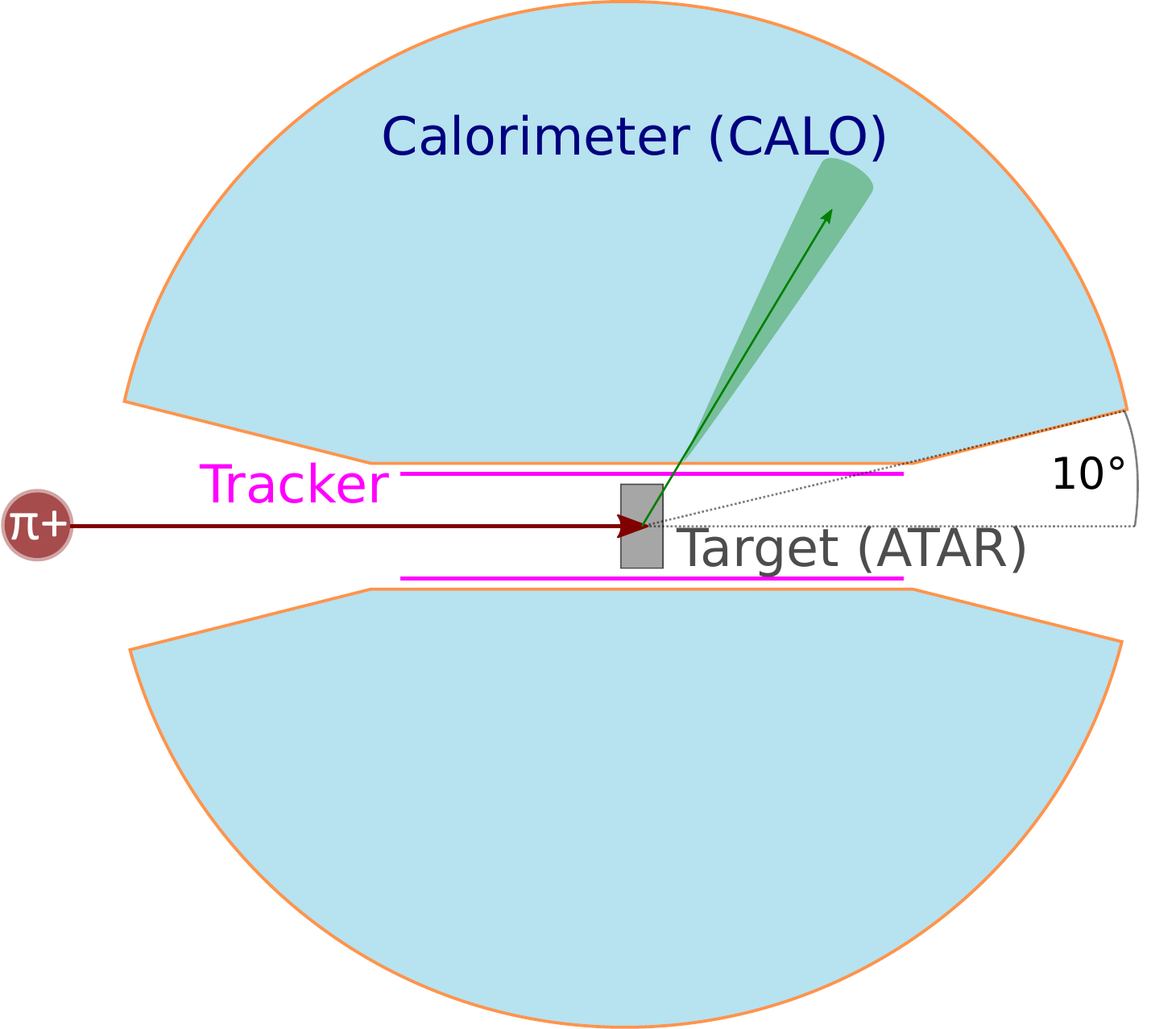}
\caption{Layout of the PIONEER rare pion decay experiment.  The intense positive pion beam enters from the left and   is brought to rest in a highly segmented active target (ATAR).   Decay positron trajectories are measured from the ATAR to an outer electromagnetic calorimeter (CALO) through a tracker.  The CALO records the positron energy, time and location.
%with  high resolution, and a hit position with modest resolution.
}
\label{fig:GenericDetector}
\end{figure}

The main challenge in developing a next generation experiment for a high precision measurement of rare pion  decays is  accurately assessing the performance of the chosen detector technology in suppressing sources of systematic uncertainties and handling increased rates. The PIONEER detector design concept
%, described in the next sections, 
is based on
the experience gathered with the PIENU \cite{PiENu:2015seu} and PEN/PiBeta \cite{Pocanic1,PEN:2018kgj, Pocanic:2014jka} experiments.
%, which are reviewed in the Appendix.
Generically, the detector will have the features sketched out in Fig.~\ref{fig:GenericDetector}. 
An intense pion beam is brought to rest in an instrumented (active) target (ATAR) and an electromagnetic calorimeter (CALO) surrounds the stopping target. A cylindrical tracker surrounding the ATAR is used to link the locations of pions stopping in the target to showers in the calorimeter.
Features of the PIONEER approach will include improved time and energy resolutions, greatly increased calorimeter depth, high-speed detector and electronic response, large solid angle coverage, and complete event reconstruction. The proposed detector will include a 3$\pi$ sr, 25 radiation length ($X_0$) electromagnetic calorimeter, an advanced design segmented stopping target, and  beam and  positron trackers.

Phase I of \nexp\ aims to measure $R_{e/\mu}$ with precision of 0.01\%, where the uncertainty budget is equally allocated to statistics and systematics; $2\times 10^8$ $\pi^+\to e^+ \nu$ events are required.
%This Proposal is specifically focused on the $R_{e/\mu}$ measurement. 
%However, we envision extension of PIONEER in Phase II (III), the details of which will be proposed  in  future. These 
Future phases will focus on a 3-fold (10-fold) improvement in the measurement of the ultra-rare pion beta decay  process,
 $\pi^+ \to \pi^0 e^+ \nu$.
The $\pi^+ \to \pi^0 e^+ \nu$ branching ratio is $ 10^4$ times smaller than the \pie~ channel and will require running with a 100x more intense pion flux.  The event identification is more straightforward owing to the characteristic signature of the $\pi^0 \rightarrow \gamma\gamma$ decay in the calorimeter.
While the optimization of the beam properties, instrumentation, and stopping target details for the pion beta decay experiment may require replacements of some systems,  the core calorimeter, mechanics, tracker, and DAQ systems will be designed to meet the needs of both experiments with limited modifications.
%In this Overview, 
 %We emphasize that the requirements for measuring $R_{e/\mu}$ drive the \nexp~ design.
 %we also discuss the differences in the setup needed for the pion beta decay measurements.
 %and outline our specific solutions and simulation studies. %\subsubsection{ Toward a realistic design for PIONEER}
%In subsequent sections we describe the beam,
%beam instrumentation, 
%target, positron detectors, calorimeter, DAQ, and electronics aspects of the experiment. The Appendices include additional technical information and priority R\&D plans.

\begin{figure}[h!]
\centering
\includegraphics[width=0.8\textwidth]{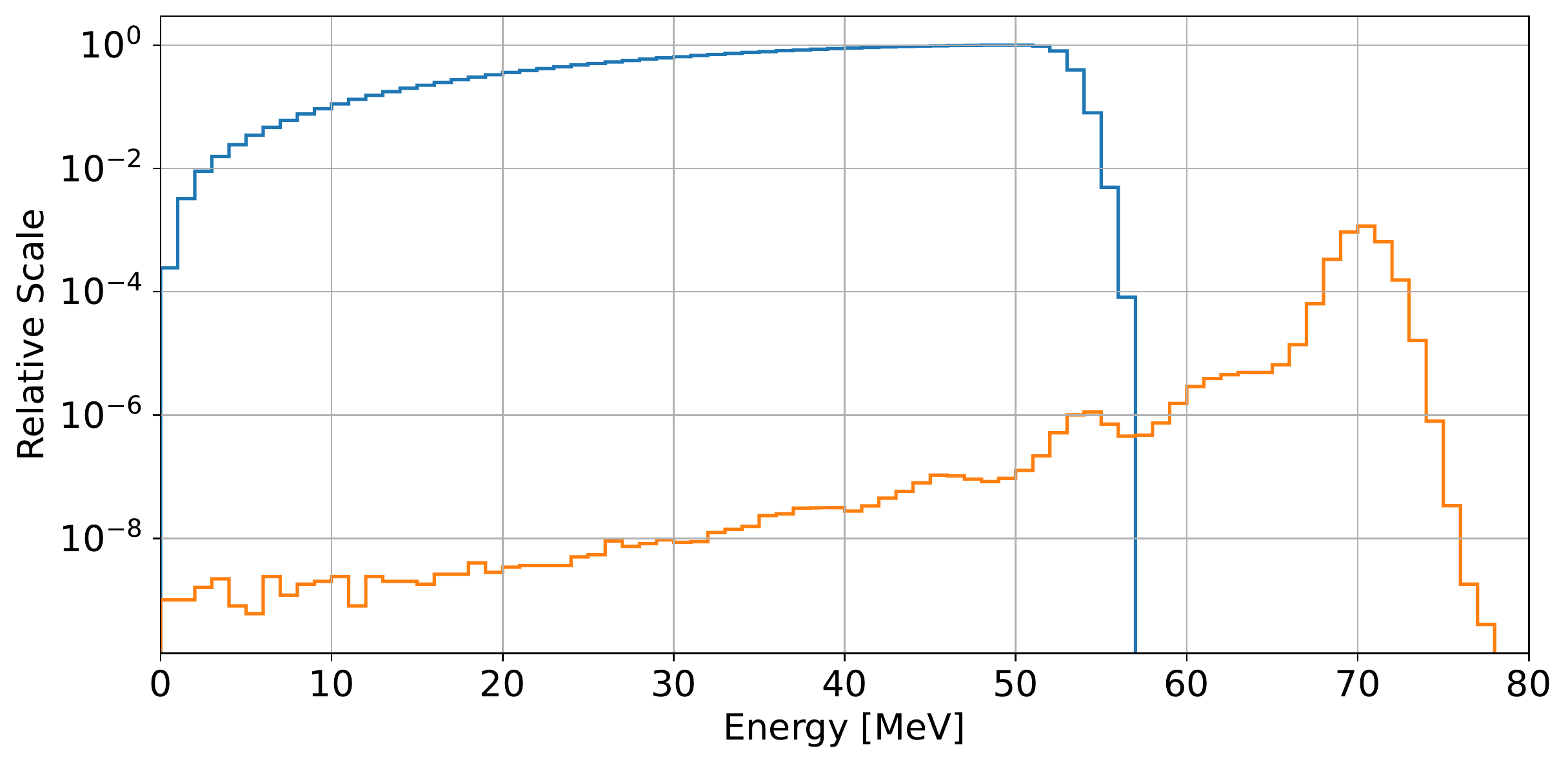}
\caption{The positron energy spectra from  muon decays (blue) and from \pie~ decays (orange) for a calorimeter 
resolution of 1.5\% and a 
depth of $25\,X_0$. 
The simulation includes energy losses owing to photonuclear interactions. 
}
\label{fig:Signal-Tail}
\end{figure}

 %\subsubsection{Requirements for measuring $R_{e/\mu}$ }
 At rest, the pion lifetime is 26\,ns and the muon lifetime is 2197\,ns.  The monoenergetic positron from $\pi \rightarrow e\nu$ has an energy of 69.3\,MeV.  Positrons from ordinary muon decay form the Michel spectrum from 0 to an endpoint of 52.3\,MeV.  In principle, the monoenergetic $e^+$ from \pie~ is well isolated above the Michel endpoint and can be easily identified using a high-resolution, hermetic calorimeter. To determine $R_{e/\mu}$ we measure the ratio of positrons emitted from \pie~ and \pme~ decays for which many systematic effects such as solid angle acceptance cancel to first order. However, counting all $\pi \rightarrow e$ events with a precision of one part in $10^4$ requires determining the low-energy tail of the electromagnetic shower and radiative decays that hide under the Michel spectrum from the $\pi \rightarrow  \mu \rightarrow e$ chain, which has  four  orders  of magnitude higher rate.

 Figure~\ref{fig:Signal-Tail} illustrates the relationship between the two channels and their respective positron energy spectra.  Here, we have modeled the spectrum from both channels assuming a high resolution, $25\,X_0$ calorimeter.  There remains an unavoidable tail fraction below 53\,MeV that must be determined accurately in order to obtain the branching ratio.
 %, and a well understood lineshape.
That challenge was critical to previous generations of  experiments and was responsible for the leading systematic uncertainty in the %most recent complement --
PIENU experiment at TRIUMF.
%and PEN at PSI -- for different, but related technical reasons. 
\nexp\enskip will minimize the intrinsic tail fraction 
%in the choice of calorimeter design.  As described below, the leading contender is a 
through the use of a $25\,X_0$ LXe calorimeter, the design of which is based on the considerable experience of the MEG Collaboration\cite{Mihara:2011zza}.
%, will be ideal.  

%The \pie~ branching ratio $R_{e/\mu}$ will be obtained by first separating events into high- and low-energy regions at an energy cut value ($E_{cut}$). % as discussed in the Appendices  for the PIENU experiment. 
%The time spectra will be fit in each region with the $\pi^+ \rightarrow e^+ \nu$ and $\pi^+ \rightarrow \mu^+ \rightarrow e^+$ timing distribution shapes, along with  backgrounds originating from different sources including event pile-up effects,  pion decays in flight, and effects from old muon decays.

%\nexp~ will incorporate improvements to the previous techniques including a deeper calorimeter.
% alone is not enough.   
%However, 
It is also important to be able to create triggers that can isolate $\pi \rightarrow e$ from $\pi \rightarrow \mu \rightarrow e$ chains within the stopping target, identify pion and muon decays in flight, as well as identify pileup from long-lived muons remaining in the target from earlier pion stops.
%Figure~\ref{fig:EventTypes} illustrates several key processes that will occur at different rates in the stopping target.  The signal event (1) has a pion decay at rest; its Bragg peak energy deposition as well as its depth within the target identifies it as a pion stop. The much higher background (and normalization) channel (2) is illustrated by that same pion stop followed by a decay to a 4.1\,MeV muon, which travels through a number of planes and then stops, leaving behind an image of its own short trajectory and Bragg peak. If the muon decays in the prompt window being observed for the pion decays (typically 3 -- 50\,ns after the pion stop), its Michel positron will be recorded in the CALO.  More rare but subtle processes 
%may also occur.
%must be understood 
%and corrected for, 
%are illustrated in the figure.  
%These include a pion that decays in flight to a positron.
%; it is a signal event, that is not accompanied by an identified stop.
%A  pion can also decay in flight within the target to a muon before it stops (3).  
%The emitted Michel positron might then be misidentified as a stopped pion.  
%In very rare cases, the emitted 4.1 MeV muon can decay in flight before stopping (4) providing a Lorentz boost to the emitted Michel positron, which  contaminates the signal spectrum above the ordinary stopped muon endpoint.  
%Finally, backgrounds can arise from upstream decays in flight.
To distinguish event types, we will use an active target that can provide 4D tracking (at the level of 150\,$\mu$m in space  and $<$1\,ns in time) and energy measurements from the O(30)~keV  signals for positrons to the 4000\,keV Bragg peaks of stopping pions and muons.  
%As discussed in Sec.~\ref{sec:ATAR}, our collaboration is focusing on the new low gain avalanche detector (LGAD) sensors as a centerpiece of the experiment.  Simulations using optimized LGAD parameters provide confidence that triggers can be constructed to isolate and measure all event types. 
New low gain avalanche detector (LGAD) sensors have been identified as meeting these requirements. Simulation studies show that, by combining information from the active target and the calorimeter, $\mu-e$ backgrounds in the tail region can be suppressed to a level 
%that will allow the 
such that uncertainty in the tail fraction contributes $<0.01\%$ to the error in $R_{e/\mu}$.

In Phase II (III), pion beta decay $\pi^+\to\pi^0 e^+ \nu$ will be measured by observing the characteristic (nearly) back-to-back gammas from $\pi^0$ decay normalized to \pie~ decay as in \cite{Pocanic5,Frlez:2003vg,Pocanic:2003pf,Frlez:2003pe,Bychkov:2008ws}. In \nexp, it is also possible to observe the low-energy positron absorbed in the ATAR in coincidence with the gammas in the calorimeter.
The Phase II (III) pion beta decay experiment will require $7\times 10^5$ ($7\times 10^6$) events at an intrinsic branching ratio of $ 10^{-8}$.  This will require running at a significantly higher pion flux of $\geq 10$\,MHz.  The beam momentum and emittance may  be higher than for the \pie~ measurement to achieve the higher flux.  
\subsection{Conceptual Design}
\begin{figure}[htbp]
	\centering
	\includegraphics[width=0.7\textwidth]{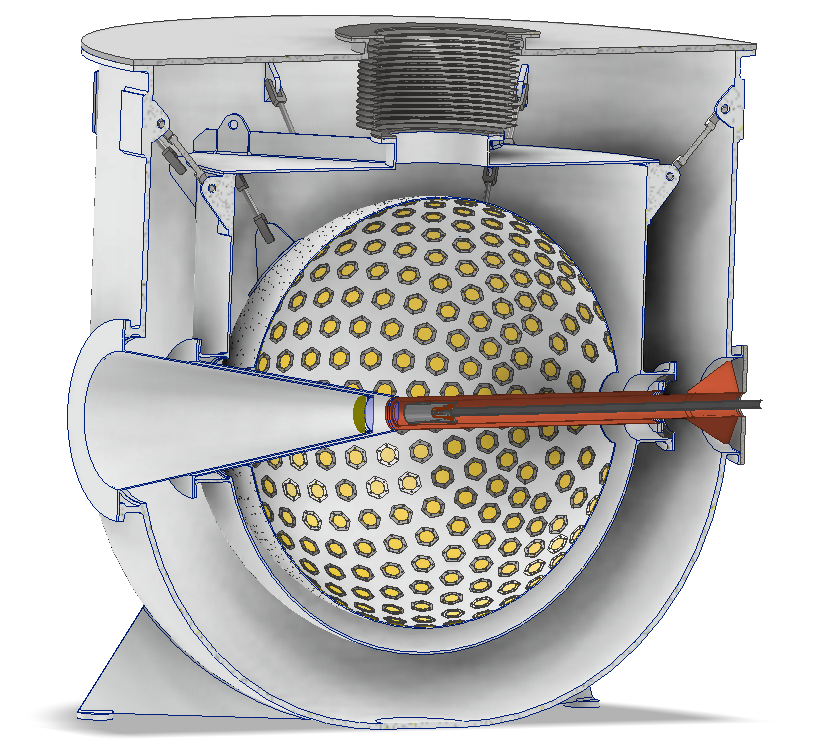}
	\caption{PIONEER conceptual design. The sphere is a LXe calorimeter, with the beam entering from the left, and the ATAR at the center. For scale, the lid is 3.05 m diameter. The yellow circles are merely representative of the LXe photosensors; they are not placed accurately.}
	\label{fig:LXe_Calo_XSec}
\end{figure}

The previous section described a generic PIONEER design in Fig.~\ref{fig:GenericDetector} and described the motivation for certain design choices. Figure~\ref{fig:LXe_Calo_XSec} shows a more specific conceptual design, which includes the beamline and beamline instrumentation, the ATAR, a cylindrial tracker surrounding the ATAR, and a liquid xenon (LXe) calorimeter. An alternative calorimeter based on LYSO cystals, which provides
natural segmentation, is also being investigated as an alternative to LXe. 
In this section, each component of this conceptual design is briefly described and opportunities for detector innovation and R\&D are identifed. More detail is available in \cite{pioneer_proposal}.

\subsubsection{Beam}
The experiment will require a continuous wave low-momentum pion beam that can be focused to a small spot size and  stop within the ATAR dimensions.  Ideal characteristics include a relatively low momentum of $55\,$\,MeV/$c$ ($\pm2\%$) and a flux of 300\,kHz.  At this low momentum, a separator is very effective to reduce background from beamline muons and positrons. The required beam properties are summarized in Table~\ref{tab:beam} and can be provided by the $\pi$E5 beam at PSI, which is a high-intensity low-energy muon and pion beamline, with a maximum momentum of 120\,MeV/$c$, that is used  for particle physics experiments, and is currently home to the MEG~II and Mu3e experiments.
Simulations for both pion production and transport of particles down the $\pi$E5 channel were performed using the \gbl\ toolkit ~\cite{Roberts:2008zzc}. Pion rates calculated in the simulation at the center of the calorimeter, for a 2\% momentum bite, 
are sufficient for the \remu\ measurement planned in Phase I of PIONEER, even when
further losses due to collimation necessary for reducing background events in the \atar\ and calorimeter are included.
For PIONEER Phase II, III increasing the momentum and the momentum bite will be required.
%can  provide the needed flux.  
%We are also investigating use of the $\pi$E1 line.
%Detector systems may be employed to image and trigger the arrival of pions into the ATAR. 
%Because of the high data rate, state-of-the-art triggering, fast digitizing electronics, and high bandwidth data acquisition systems are required.

\begin{table}[htb]
\centering
\begin{tabular}{ccccccc}
\toprule
Phase & p  & $\Delta$p/p & $\Delta$Z 	& $\Delta$X x $\Delta$Y& $\Delta$X',$\Delta$Y'  & R$_\pi$ \\ 
      & (MeV/c) & (\%) & (mm) 				& (mm$^2)$ 				& 						 & (10$^6$/s) \\ 
\hline 
I & 55-70 & 2 & 1 & 10x10 &  $\pm$10\degree & 0.3 \\ 
II,III & \app\ 85 & $\le$ 5 & 3 & 15x15 & $\pm$10\degree  & 20 \\ 
\bottomrule 
\end{tabular} 
\caption{Required beam properties. $\Delta$Z and $\Delta$X $\times$ $\Delta$Y are longitudinal (FWHM) range width and transverse (FWHM) beam sizes at target location, respectively.}
\label{tab:beam}
\vspace{-0mm}
\end{table}

\subsubsection{Active Target}
\begin{figure}[htbp]
\centering
% \includegraphics[width=0.8\textwidth]{Figures/event_display.png}
%\vspace{-20mm}
\includegraphics[width=0.75\textwidth]{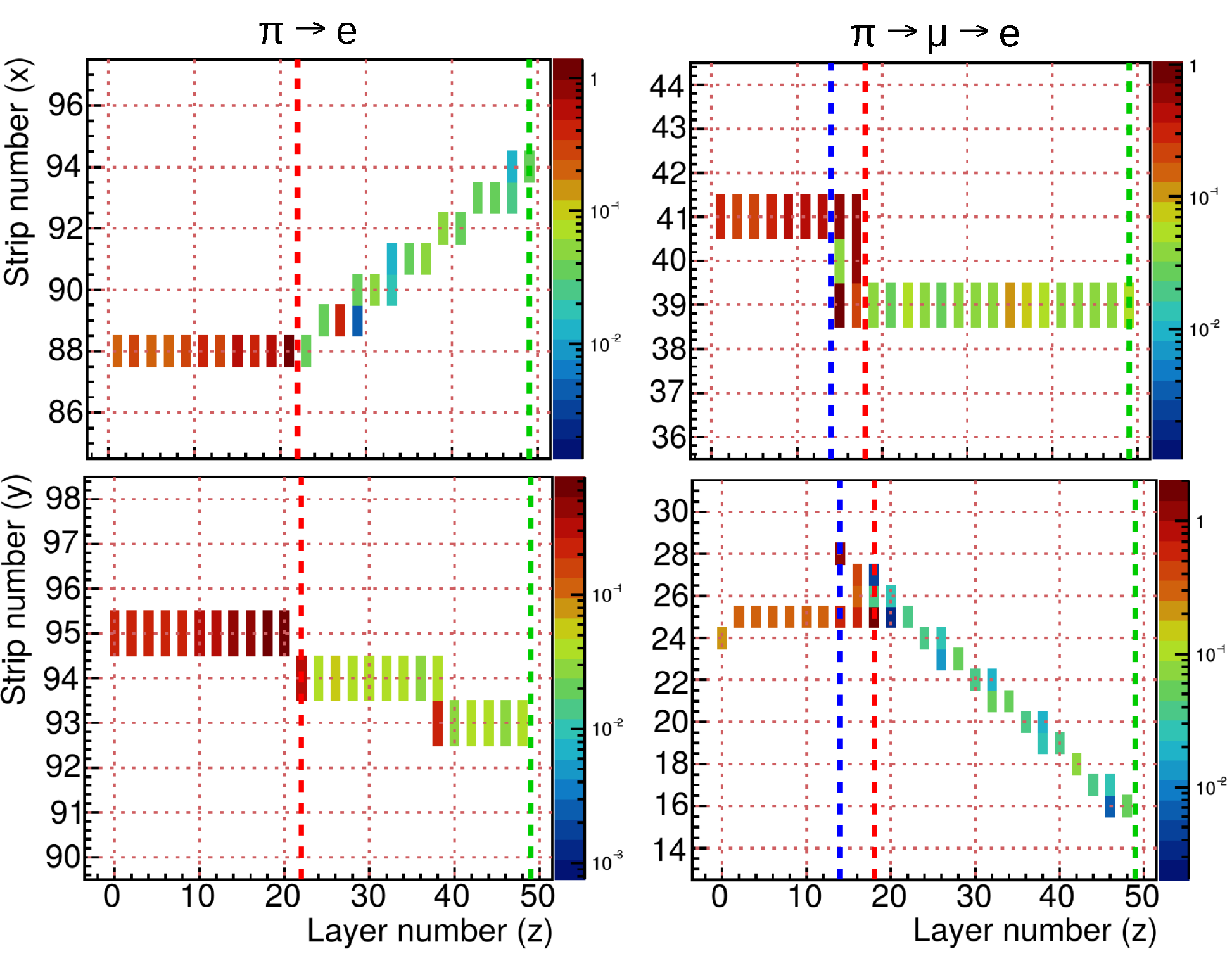}
\caption{Simulated example displays of pion decay events in the \atar. Pions enter horizontally from the left; the  red dotted lines show the positions of the pion stops.
%and the green dotted lines show the position of positron exiting the ATAR.
The color of the bars indicates the deposited energy. Left column: X-Z (top) and Y-Z (bottom) strip  views of a \pie\ event.  Right column: Same views of a \pme\ event. The  blue dotted line shows the position of the decay muon stop.}
\label{fig:Event_display}
\end{figure}

A highly segmented active target (\atar)~\cite{Mazza:2021adt} is a key new feature of the proposed PIONEER experiment. Relative to previous experiments, the use of \atar\ allows for discrimination against backgrounds by looking for deposition patterns internal to the target that are associated with the various signals' decay topologies. The \atar\ will define the fiducial pion stop region, provide high resolution timing information, and will furnish selective event triggers. Examples of simulated \atar\ event displays for \pie\ and \pme\ events are shown in Fig.~\ref{fig:Event_display}.

To adequately supress background from $\pi \rightarrow \mu \rightarrow e$ decay at rest (\pdar), pion and muon decay-in-flight (DIF), and accidental muon stops that precede the trigger signal, the \atar\ must be able to detect both the exiting $e^+$, a minimum ionizing particle (MIP), and larger ($\thicksim$100 MIPs) energy deposits from $\pi^+$ and $\mu^+$.
The large dynamic range O(2000) of the signals is a significant challenge for the readout electronics  in the amplification and digitization stages.
The position of the energy deposits needs to be identified with sufficient granularity along the beam direction and in the transverse plane.
Furthermore, to identify single components of the decay processes a \SI{\sim 1.5}{\ns} pulse pair resolution is needed.

The chosen technology for the \atar\ is Low Gain Avalanche Detector (LGAD)~\cite{bib:LGAD}, 
thin silicon detectors with moderate internal gain; the LGAD technology was chosen over standard silicon technology because of the intrinsic gain and thin bulk.
A \SI{120}{\micro\metre} thick LGAD sensor, coupled to fast electronics, has a time resolution of less than \SI{100}{\ps} on the rising edge and can separate a single hit from two overlapping hits if they arrive more than \SI{1.5}{\ns} apart. There are two specific LGAD technology options under
consideration. The first option, AC-LGADs, overcome the granularity limitation of traditional LGADs and have been shown to provide spatial resolution of the order of tens of $\mu$m~\cite{Tornago:2020otn}. AC-LGAD design also allows to have a completely active sensor with no dead regions. The second option is Trench Isolated (TI) LGADs, which are a novel silicon sensor technology that utilizes a deep narrow trench to electrically isolate neighboring pixels to prevent breakdown, as opposed to standard LGADs which use a junction termination extension to prevent breakdown at the pixel edges \cite{9081916}. By utilizing the deep trench isolation technology, the no-gain region is reduced to a few micrometers, thus achieving a higher fill factor than regular LGADs.

\begin{figure}[htbp]
\centering
\includegraphics[width=0.4\textwidth]{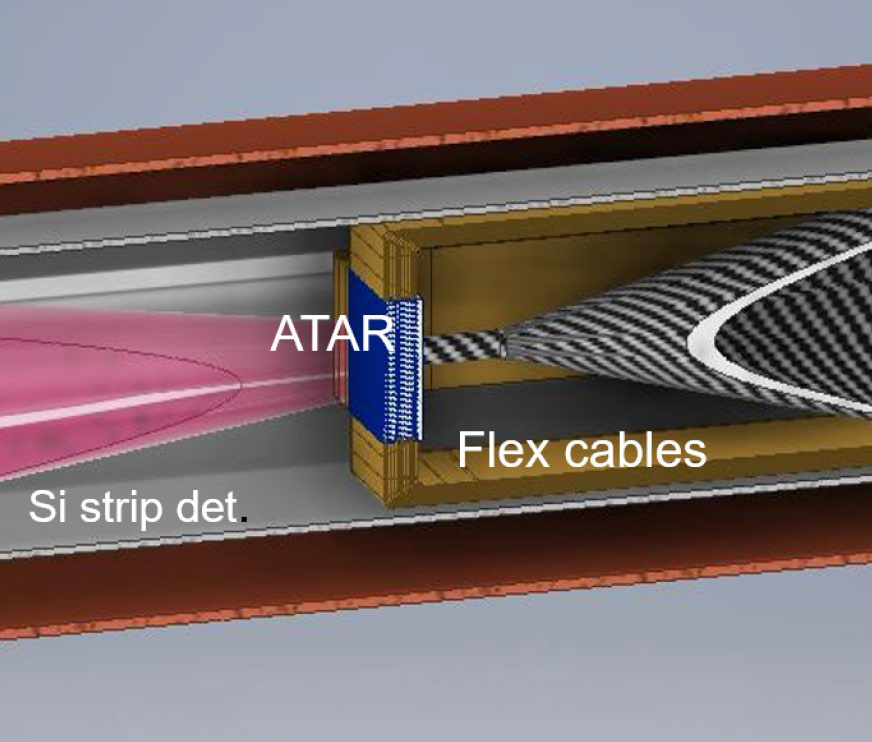}
\includegraphics[width=0.55\textwidth]{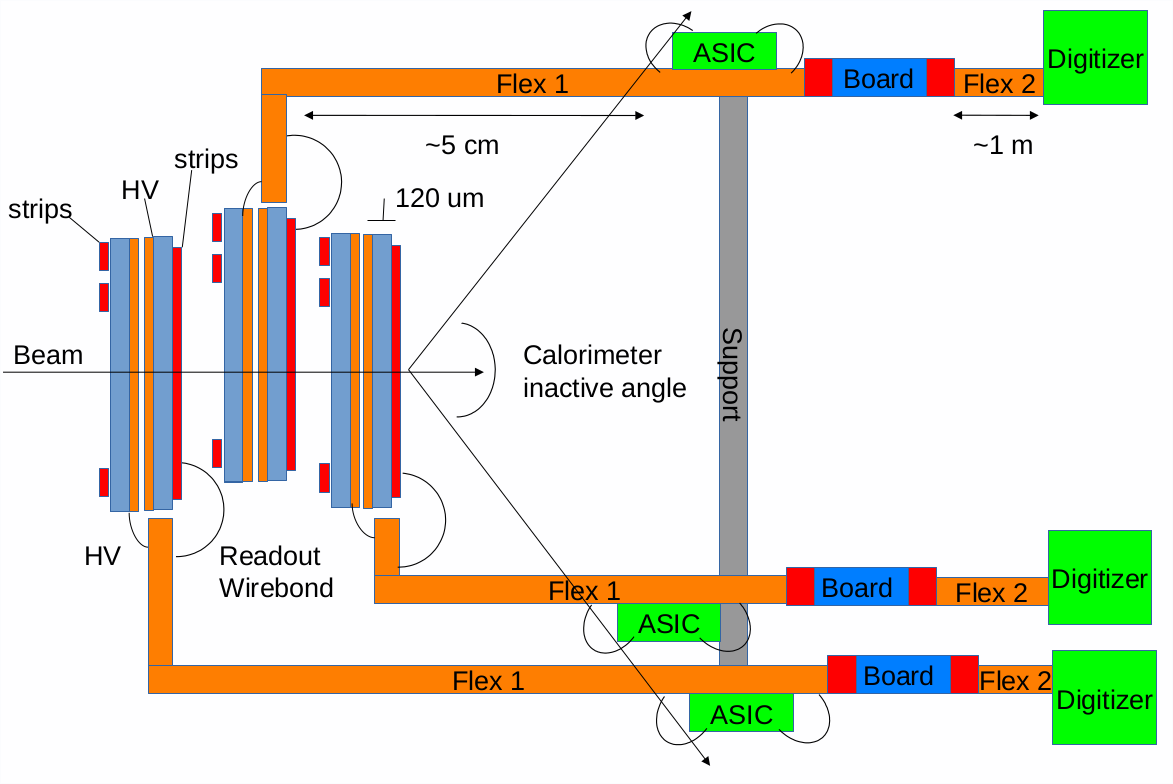}
\caption{Left: \atar\ position in the beam line. Right: Concept schematic design of the \atar. The flex from the first, third and fifth sensors is directed in and out of the page. The modules are attached on the HV side and with a few  \si{\micro\metre} of separation on the strip side. 48 sensors are coupled in 24 pairs. }
\label{fig:ATAR_scheme}
\end{figure}

A preliminary design for the \atar\ is shown in Fig.~\ref{fig:ATAR_scheme}. In this design, the \atar\ dimensions are 2$\times$2\,\si{\cm\squared} transverse to the beam. In the beam direction individual silicon sensors are tightly stacked with a total thickness of \SI{\sim 6}{\mm}.
The detector is arranged in a strip geometry, with the strips oriented at \ang{90} to each other in subsequent staggered planes to provide measurement of both coordinates of interest, and with the electronic readout connected on the side of the active region via wire bonding.
The sensor geometry has strips with a pitch of \SI{200}{\micro\metre}, so that a sensor would have 100 strips mated to a chip with 100 channels and \SI{2}{\cm} width, a standard dimension for microchips. Sensor thickness is around \SI{120}{\micro\metre}, such that $\thicksim$50 planes are needed to reach a total thickness of about \SI{6}{\mm}.
The detectors are paired with the high-voltage facing each other in a pair to avoid ground and high voltage in proximity and the strips are wire bonded, with a connection alternating on the four sides, to a flex that brings the signal to a readout chip positioned a few cm away from the active volume.
Readout from both ends is being investigated to reduce the average material traversed by exiting positrons; in the single-sided readout design, the maximum material in the path of the positrons occurs when 12 flexes are traversed. 
%The readout ASIC sits on the first flex that tapers out to accommodate the additional traces. 
%Then the flex is connected, via connector, to a PCB connected to a second flex that brings the amplified signal to the digitizers in the back end.

R\&D and optimization of the \atar\ design is ongoing. Preliminary LGAD studies done with X-rays coming from the Stanford light source (SSRL)~\cite{GALLOWAY20195} and PSI~\cite{Andra:ay5534} show that LGADs have an energy resolution of around \SI{10}{\percent}. The effect of gain saturation that was reported in the community in the past year~\cite{gainsuppr} also needs to be studied.
Strip AC-LGAD prototypes from Brookhaven National Lab (BNL) have been tested with a laboratory IR laser TCT station~\cite{Particulars} and at a Fermilab (FNAL) test beam~\cite{ACLGADpico}; the position resolution of this prototype sensor varies between 5-15~$\mu m$ in the direction perpendicular to the strip.
Prototypes TI-LGADs sensors from Fondazione Bruno Kessler (FBK) \cite{9081916} were studied at UCSC and found to have the standard response of a conventional LGAD with a small amount of “cross-talk” constant response of the sensor along the strip. In the next few years, several prototypes will be tested in laboratories and test beams to identify an optimized LGAD configuration for PIONEER.

The amplifier ASIC to read out the \atar\ needs to be fast enough for the sensor in use; for the signal rise time in the 120\,$\mu$m-thick prototype sensors,  a bandwidth of 1\,GHz should be sufficient.
However, the high dynamic range (2000) requirement for the ATAR brings major complications to the readout.
Current fast readout chips usually have a dynamic range of $<$~1000, since they are targeted at MIPs-only detection in tracker sub-systems.
One possibility is to develop an amplifier chip with logarithmic response or dynamic gain switching as well as a high enough bandwidth, currently no such chip exists with the necessary characteristics.
Already available integrated chips, such as FAST~\cite{OLAVE2021164615} and FAST2, are being evaluated.
Some new ASIC technologies that are being developed at UCSC in collaboration with external companies can run with 2.5V maximum signal, this allows for an increased dynamic range.

To successfully reconstruct the decay chains, the ATAR is expected to be fully digitized at each event.
To achieve this goal, a high bandwidth digitizer with sufficient bandwidth and sampling rate have to be identified. 
The same issue afflicting the amplifier, the high dynamic range, is also problematic for the digitization stage. 
A digitizer that would suit PIONEER's requirements needs to be identified; a ready commercial solution would be the best option but the cost per channel might be prohibitive. 
For this reason the collaboration is exploring the possibility to develop a new kind of digitizer specific to this application.

\subsubsection{Cylindrical Tracker}
A dual layer cylindrical silicon strip tracker is situated between the \atar\ and the calorimeter to measure the  positron position in two dimensions (along the beam direction, $z$, and azimuthal angle, $\phi$), and time.
The detector has  an inner diameter of \SI{5}{\cm} and a length of \SI{25}{\cm}. Overlapping lengths of long strips (about \SI{10}{\cm}) are needed to cover the entire region. Two layers of strips with a small stereo angle  between them will provide 
O(mm) $z$ resolution and \SI{300}{\micro\metre} resolution in the direction perpendicular to the strips. An alternative under consideration is to connect  two or three strip sensors in a line reading out both ends to obtain O(mm) position information along the strip position using charge attenuation information. The silicon strip sensors may be constructed with either regular silicon or LGADs.

\subsubsection{Liquid Xenon Calorimeter}

Due to its fast timing properties, high light yield with excellent energy resolution and highly uniform response, liquid xenon (LXe) 
read out by UV sensitive phototubes and state-of-the-art vacuum ultraviolet (VUV) silicon photomultiplier (SiPMs) is considered for the calorimeter. Here, experience is drawn from the  MEG \cite{MEG:2016leq} and  MEG-II \cite{MEG:2018vi} experiments which use a large scale,  high rate  LXe detector to search for the lepton flavor violating muon decay, $\mu^+\rightarrow e^+ \gamma$.
 Experiments searching for elusive dark matter (e.g. XENON, LUX-ZEPLIN) and (neutrinoless) double beta decay events (KamLAND-Zen, (n)EXO) also use detectors with similar scale liquid xenon cryostats. PIONEER, like MEG, detects only scintillation light (other experiments rely on both scintillation and charge collection) and is a high rate experiment. The PIONEER LXe detector is foreseen to be a 25 radiation length, 3$\pi$-sr sphere surrounding the ATAR. 
 
 The homogeneity of the LXe detector is an advantage in achieving the high energy resolution which is important for determining accurately the low energy ``tail'' fraction of $\pi\rightarrow e \nu$ events. MEG  currently reports  an energy resolution of $\sigma=1.8\%$  for 50\,MeV gammas and they continue to study possible improvements. The baseline energy resolution goal for PIONEER at 70\,MeV is 1.5\%.
 
 The  energy resolution is impacted by the efficiency of collection of scintillation light which is itself influenced by design parameters (like photo-sensor coverage) and physical or technical parameters (like the light attenuation due to impurities in LXe, reflection of VUV light on surfaces, photo sensor refraction index, the level of dark current which impact the photo-electron threshold for summing the energies of the different photo-sensors, etc). In addition to the finite energy resolution of the calorimeter, photo-nuclear interactions, shower leakage and geometrical acceptance contribute to the low energy tail. The impact of photo-nuclear effects in LXe, for which little literature exists, will be determined by simulation and bench-marked against available data and new measurements. 
 
\begin{figure}[h!]
\centering
\includegraphics[width=0.9\textwidth]{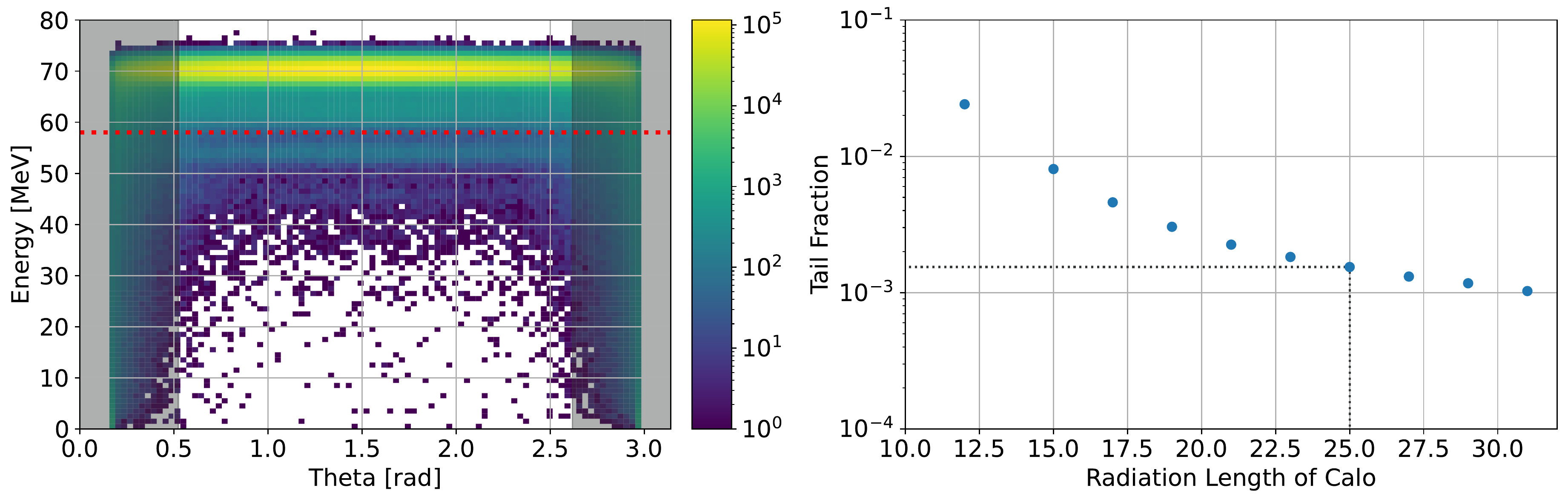}
\caption{Left:  The energy deposited by monoenergetic 69.3\,MeV $e^{+}$ from \pie~ decays in a 25\,$X_0$ calorimeter with an energy resolution of 1.5\% vs. the angle Theta with respect to the beam axis.  The grey bands indicate the boundaries of the fiducial volume region (here, $30^{\circ}$).  Right: The shower tail fraction below 58\,MeV vs. the calorimeter depth in radiation lengths for the 69.3\,MeV $\pi \rightarrow e \nu$ events. 
%This does not include contributions from pion radiative decays.
%resolution of 1.5\% and a depth of $25\,X_0$.
}
\label{fig:tail_fraction}
\end{figure}

A GEANT-4 simulation of a bare LXe calorimeter geometry was used to determine the residual tail fraction below the Michel end point versus calorimeter depth for \pie~ events.  Figure~\ref{fig:tail_fraction}, Left, shows the energy deposited in the spherical calorimeter vs. the angle Theta with respect to the beam axis for a 25\,$X_0$ calorimeter depth.  Figure~\ref{fig:tail_fraction}, Right, shows the fraction of the energy deposited that is below  58\,MeV  vs. depth.  The volume of LXe required scales as the radius cubed and the required photo-sensor coverage scales as the radius squared.  These practical factors are optimized for smaller depth.  The containment of the shower slowly improves with increased depth. %, but it is a slow function.  
%The depth was chosen to be 25\,$X_0$, which we use throughout this proposal.   
This choice, along with the spherical shape and %$>\unit[3\pi]{sr}$ coverage, limits the effect of secondaries being missed by the calorimeter (e.g.~Bhabha scattered events) and will significantly reduce the contribution of shower leakage. We will continue to carefully 
optimization of the calorimeter geometrical parameters and assessment of the expected detector sensitivity is ongoing.

In order to readout the liquid xenon scintillation light, special photo sensors are necessary due to the short scintillation wavelength of 175\,nm and the low operation temperature of 165\,K. Two types of photo sensors were developed for this purpose by Hamamatsu K.K. and the MEG/MEG II collaboration: 2-inch photomultipliers (PMT) and multi-pixel photo sensors (MPPC).
A round-shaped 2-inch PMT sensitive to VUV light (R9869) 
%was developed using
%Bialkali  (K-Cs-Sb) photocathode and a synthetic silica window. It 
achieved a quantum efficiency of \SI{15}{\percent}. 
The gain of the PMTs was found to decrease due to damage to the dynode induced by photoelectrons; however the PMT lifetime is sufficiently long to not be of concern for PIONEER.
%Aluminum strips were added in the photocathode to prevent an increase of the sheet resistance at low temperature.
In order to improve the granularity for the inner surface in the MEG-II experiment, a silicon photomultiplier, 
%called MPPC (S10943-4372) 
with an active area of 12$\times$12\,mm$^{2}$, was developed. 
%The main developments consisted in removing the protective layer, matching the optical properties of the silicon surface to that of the LXe, and reducing the thickness of the insensitive layer. 
%Besides achieving higher granularity, 
There are several advantages, in additional to higher granularity, to using SiPMs with respect to photomultiplier tubes: they are insensitive to magnetic fields, the single photoelectron peak can be used for calibration of the sensor, and the required supply voltage is relatively low (less than \SI{100}{\volt}). However, while the MEG-II collaboration reported a photon detection efficiency (PDE) of $\sim20\%$ \cite{MEG:2018vi}, they observed later degradation of the PDE in LXe \cite{meg2sipm} which is under investigation. For this reason, the baseline design for PIONEER is the use of PMTs on the outer surface targeting a coverage of \SI{20}{\percent} (1000~phototubes) of the surface. The choice of photosensor technology may evolve depending on the developments regarding SiPM performance degradation.

\begin{figure}[tb]
\begin{center}
\includegraphics[width=0.75\textwidth]{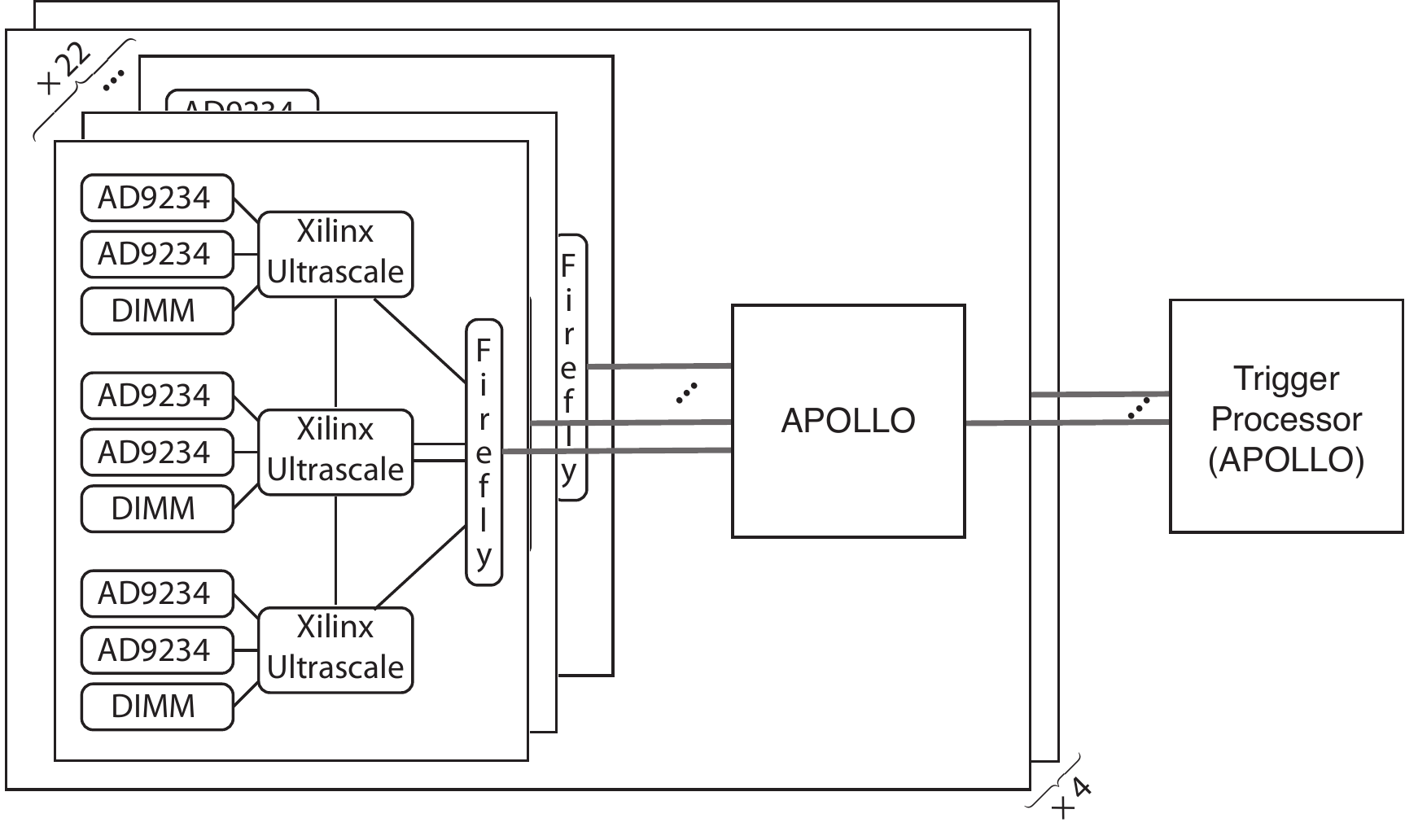}
\caption{Proposed calorimeter digitization and readout.  The 12 channel digitizer boards utlize the dual channel AD9234 12 bit, 1 GSPS ADCs, Xilinx Ultrascale FPGAs for contro, and Samtec Firefly\textsuperscript{TM} high speed communications to a CMS APOLLO board.  The APOLLO board can receive up to 22 boards in this configuration -- instrumenting quadrant of an order 1000 channel calorimeter.}
\label{fig:calodaq}
\end{center}
\end{figure}

The digitization and readout electronics proposed for the calorimeter are shown in Fig.~\ref{fig:calodaq}. The 12 channel digitization boards will utilize the Analog Devices AD 9234 dual channel, 12 bit 1 giga-samples per second (GSPS) ADC chip, chosen for its low latency of 59 ns from presentation of the signal at the front end to output of the digitized signal. Calorimeter information can be summed with a pipelined adder and potentially corrected with a running pedestal measurement as the first stage in a total energy measurement.  By clocking the ADC at the slightly lower rate of 976\,MHz, the ADC information can be synchronized to the firefly data transfer rate of 244\,MHz, simplifying synchronization of the system. The ADCs will sample continuously with samples stored initially in a ring buffer on the FPGA.  Upon receipt of a trigger, a configurable time window will be stored in DIMM memory.  By deploying a DIMM with a 128 bit data path, simultaneous reading and writing can be accomplished via two 64 bit pathways. A single FPGA will control one pair of ADCs (four calorimeter channels).  It can compare the energy sum from each channel against a channel activity threshold (or thresholds), as well as combine the four running energy sums as the first stage of a total energy sum for the high energy trigger.  That FPGA will also drive a single firefly channel. The digitizer boards will communicate with intermediate Apollo boards via the 16.1\,GHz firefly links, which come packaged with a minimum of four individual links.  Three of these links will provide the TCDS (or equivalent) clock and control signals and send the trigger and channel readout information to each of three sets of 4 ADC channels.  The fourth firefly line will provide PCIe or Ethernet communication to allow board configuration and other slow control functions.

The xenon scintillation light absorption length has been  measured for MEG to be more than \unit[1]{m} \cite{Baldini:2004ph}.
 %which is highly advantageous when targeting a high energy resolution as is needed for PIONEER. 
 However, the light absorption length can be significantly reduced through absorption by impurities such as H$_2$O and O$_2$.
A large scale detector requires purification at the ppb level  which was achieved by MEG.
The purification system and the cryostat design (needed to maintain the xenon at \unit[165]{K} \cite{Mihara:2011zza}) of MEG will be considered for scaling up the design for PIONEER. Details of the cryogenics, purification, and mechanical engineering have been considered and are described in more detail in \cite{pioneer_proposal}.
%(see Sec.~\ref{sec:LXe_Cryogenics-purification}).
%, see section \ref{sec: LXe_photosensors}.

\begin{figure}[h!]
\centering
\includegraphics[width=\linewidth]{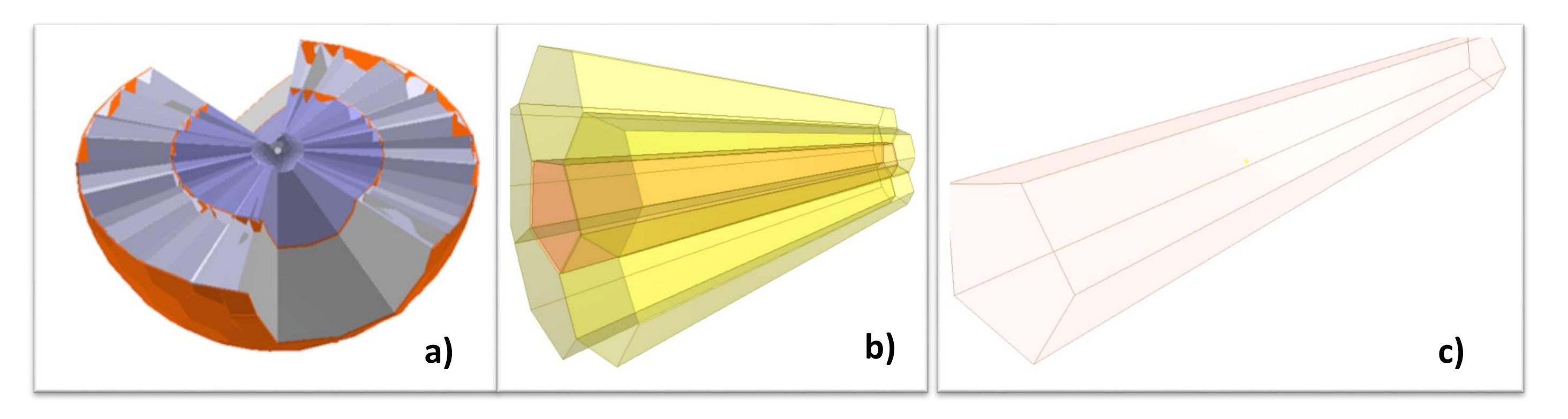}
\caption{Possible use of an inner array of tapered LYSO crystals within the open volume of the existing PEN CsI calorimeter.  a) Opened view showing in blue the array of LYSO crystals that matches one-to-one to the existing geometry of the PEN crystals shaded in gray.  b) An example array ideal for testing the concept.  c) An individual pentagonal crystal, 16$X_0$ in depth.  Each such crystal would be read out by a thin array of SiPMs.
}
\label{fig:LYSO-figs}
\end{figure}

\subsubsection{Considerations on the Alternative LYSO Calorimeter}
A naturally segmented array of tapered LYSO crystals provides an attractive alternative to our proposed LXe-based calorimeter.  
As shown in Fig.~\ref{fig:LYSO-figs}a, we are exploring a geometry that matches that of the PEN pure CsI detector~\cite{Pocanic1,Frlez:2003vg,Pocanic:2003pf,Frlez:2003pe,Bychkov:2008ws}. 
An inner array of LYSO crystals, a Cerium doped Lutetium based scintillator, 
can be inserted between the ATAR and the existing PEN CsI array. 
Figures\,\ref{fig:LYSO-figs}b and c indicated prototype geometries that are designed to fit inside the PEN calorimeter.
On this option, we have collected pros and cons as well as experiences of other groups from experts in the field~\cite{Zhu:2021sij}.
%The PEN crystals were limited to 12$X_0$  depth, which is insufficient for an ideal $\pi \rightarrow e \nu$ measurement.  However, the segmentation and fast response allows for various trigger patterns and it is well-designed for the pion beta decay phase in the PiBeta configuration. This detector has an inner radius of 26\,cm, and an outer radius of 48\,cm.  
%CsI has a radiation length of 1.86\,cm. With the compact ATAR geometry we are proposing, a sufficient volume exists to insert an inner array of crystals between the ATAR and the existing PEN CsI array, see Fig.~\ref{fig:LYSO-figs}a.   On paper, LYSO crystals appear to be the ideal choice for such an array.  
LYSO is radiation hard, non-hygroscopic, and has high density ($X_0$= 1.14\,cm, $R_{M}$=2.07 cm) and a light yield comparable to the highly luminous NaI (Tl), but with much faster light signals.  Its 420\,nm typical scintillation light has a 40\,ns single exponential decay time and the spectrum is well matched to conventional SiPM photosensors.  On the other hand, the growth of relatively long LYSO crystals is a fairly new and expensive R\&D effort, and the energy resolution may be a limitation based on existing tests~\cite{Atanov:2016blu}. In order to advance this alternative design, it is imperative to improve on the uniformity of light production and transmission along the length of the crystal~\cite{Mao:2012dr}. We plan 
to perform R\&D regarding the possibilities of using LYSO for PIONEER. 

\subsubsection{Trigger and Data Acquisition System}

\begin{figure}[htb]
\centering
\includegraphics[width=0.7\textwidth]{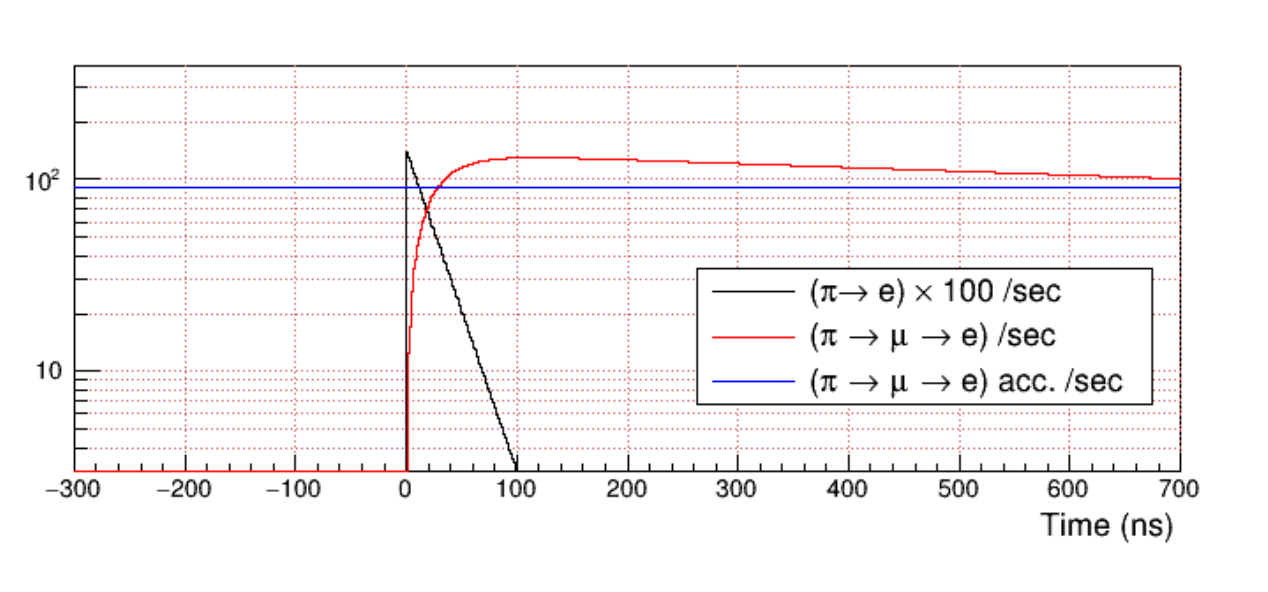}
\caption{Positron rates after PI per second in 1\,ns time bins. \pie\ positron rates are multiplied by a factor 100.  \pme\ rates generated by PI shown in red and positron rates from old muons, i.e. from accidental pions, shown in blue. }
\label{fig:time}
\end{figure}

All triggers will start with a PI signal, which is a loose condition for an incident beam particle defined as a coincidence of the beam detectors upstream of the ATAR.
%(LGAD telescope layers). 
The key point is that this trigger must not introduce any bias between \pie\ and \pme\ events. 
%For the estimates we assume that a separator is installed, so that the rate R$_{PI}$=300\,kHz corresponds to the design beam rate and is dominated by pions.
The main time distributions in the vicinity of the PI signal are sketched in Fig.~\ref{fig:time}. After an initial build-up with the pion lifetime,  positron rates from \pme\ reach their maximum before decreasing with the muon lifetime. The constant accidental rate from muon decays of other pions stopped in ATAR is high.

After requiring the PI signal, several main triggers are formed, including a minimum bias trigger, with the PI signal prescaled by about k=1000, a trigger to select high energy (E$_{tot} \gtrsim$ \thcal) events detected by the \calo\ within a time range TR=[-300,700] ns relative to PI, a trigger for all events with a \track\ hit within time range TR relative to PI, prescaled by about k=50, and a trigger to 
select prompt events with a \track\ hit in time range [2,32] ns relative to PI, potentially prescaled. For all triggers, a full event readout, including
the \calo\ waveforms, the \atar\ and the \beam\ and \track\ detectors,
will be initiated.

\begin{figure}[H]
\begin{center}
\includegraphics[width=0.6\textwidth]{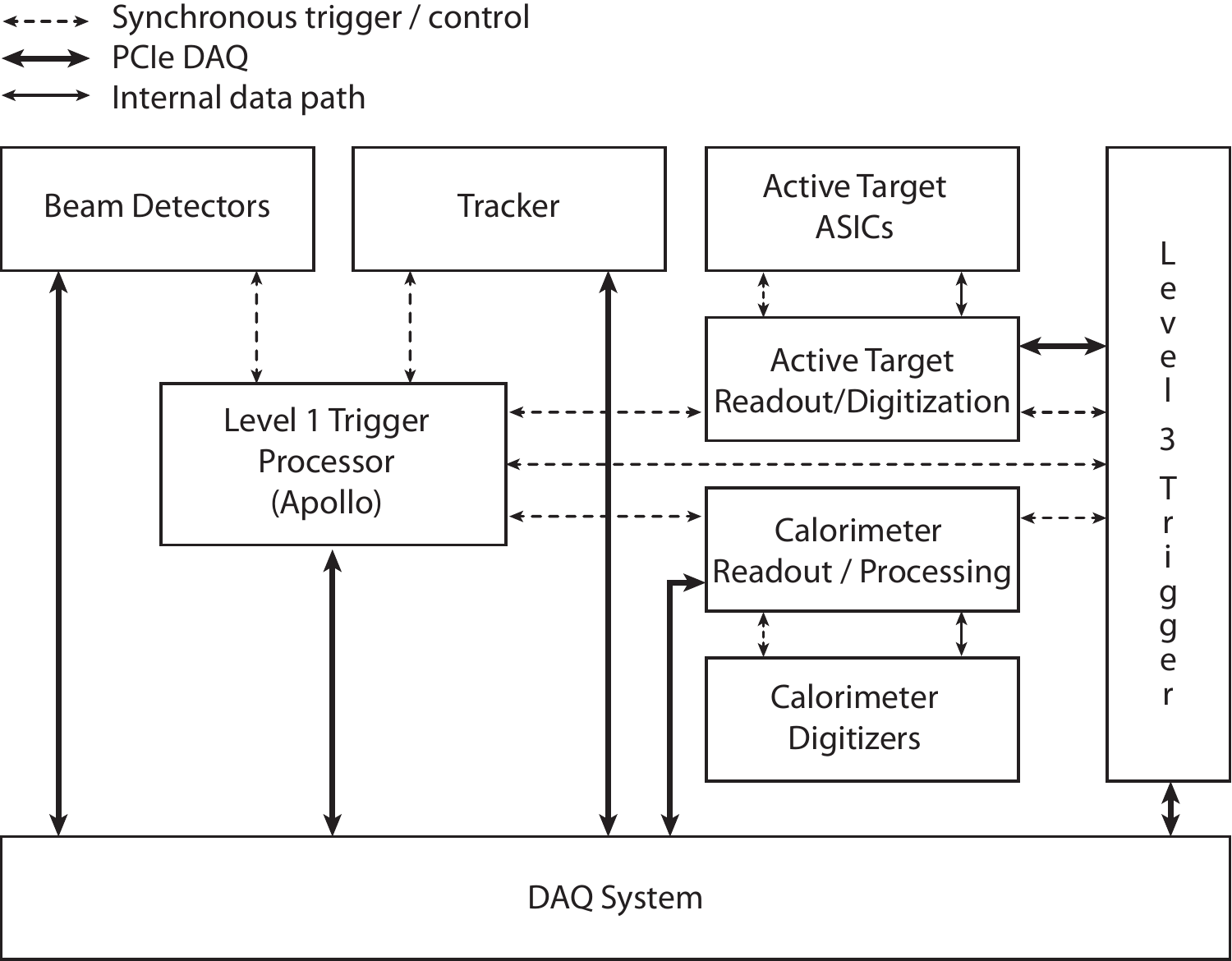}
\caption{Proposed topology of the trigger and subdetector readout systems.}
\label{fig:readout}
\end{center}
\end{figure}

Figure~\ref{fig:readout} shows the planned topology for readout and triggering of PIONEER.  The design takes advantage of
the APOLLO~\cite{ApolloTWEPP21} platform designed for the trigger track finder and pixel readout in the CMS experiment at
the LHC. The platform supports flexible high speed synchronous command and triggering, as well as options for direct IO for receiving timing signals or generating trigger signals for devices expecting analog triggers. While by some measures this board likely provides more power than needed for PIONEER, it has already undergone significant prototyping with full production anticipated in 2024, and can satisfy several of the needs of the system with a single platform.  The engineering group that designed the CM will also contribute to calorimetry readout, providing additional coherence.  

The PIONEER data acquisition system must handle the readout, event assembly and data storage for the active target, positron tracker, electromagnetic calorimeter and other detector sub-systems of the experimental setup. It must provide a deadtime-free, distortion-free record of the datasets identified by the various physics and calibration triggers. It must facilitate the monitoring needed to guarantee the overall integrity of data taking and provide the metadata needed to document the experimental configuration during data taking. Finally, it must enable the configuration of the readout electronics and the associated trigger, clock and control system.

\begin{figure}[H]
\begin{center}
\includegraphics[width=0.75\textwidth]{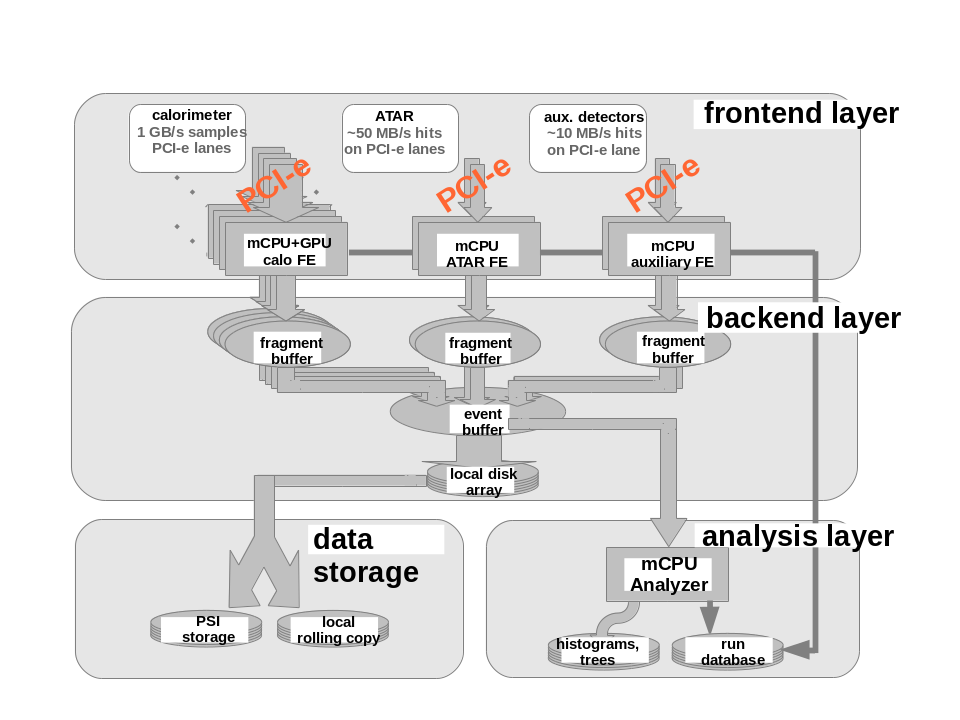}
\caption{Schematic of the data acquisition system showing the frontend layer for data readout and experiment configuration, the backend layer for event assembly and data storage, and the analysis layer for data quality monitoring. The number of frontends and the topology of the FPGA-to-frontend and frontend-to-backend networks will be based on the calorimeter, ATAR and FPGA technology choices.}
\label{f:daqlayout}
\end{center}
\end{figure}

The acquisition will be implemented as a modular, distributed system on a parallel, layered processor array consisting of networked, multi-core, commodity PC's running an operating system. The overall layout is depicted schematically  in Fig.\ \ref{f:daqlayout}.  It will comprise a frontend processor layer responsible for readout and processing of event fragments from the FPGA-based fast electronics instrumenting the various detector sub-systems, a backend layer responsible for event building and data storage, and an analysis responsible for monitoring of data integrity.

The DAQ software will be based on the MIDAS data acquisition package \cite{midas}, CUDA GPU toolkit \cite{cuda}, ROOT data analysis package \cite{root}, and Linux PCI Express system and utility libraries. The MIDAS software consists of library functions for data flow between different processes on local / remote devices as well as infrastructure for data logging, experimental configuration and local /remote run control. It also incorporates an integrated alarm system and slow control system. The devices drivers for the configuration and the readout of the Apollo board FPGAs will be based on the PCI Express communication protocols / libraries.

The DAQ will process in real-time the data from a roughly 3.5 GB/sec raw data rate to a roughly $\sim$300 MB/sec processed data rate for data storage on PSI's Petabyte archive. One option for the data processing is the lossless compression of the slow decay-time calorimeter signals via a combination of delta encoding and Golomb coding. Other possibilities are zero suppression of \calo\ islands and realtime fitting of \calo\ pulses. These algorithms are well suited to parallel processing using GPUs.

R\&D is planned to demonstrate both the technology for the FPGA-to-CPU / GPU communication via optical PCI-express links and the performance of the data compression schemes. %The R\&D setup will allow for code development, testing and debugging as well as the evaulation of the rate capabilities and the compression capabilities of the system. We plan to use commercial PCI-Express FPGA development boards as data simulators for the detector sub-systems in order to prototype and benchmark the DAQ.
The conceptual design and R\&D plan draw on experience with similar architectures of distributed data acquisition systems for the MuLan, MuCap and MuSun experiments at the PSI and the g-2 experiment at FNAL.

\subsection{Simulations}
Each of the design elements discussed above is being actively studied using GEANT4-based \cite{GEANT4:2002zbu} simulations.
The simulation efforts include 
beamline and upstream detector simulations, simulation of the active target, and simulation of the calorimeter. The $\pi$E5 beamline at PSI is simulated using G4Beamline \cite{Roberts:2008zzc}. The remainder of the simulation is done primarily using GEARS, 
an extension of GEANT4, which streamlines readout for rapidly iterating systems. The geometries for each of the experimental components are generated using a stand-alone Python script and geometry library, which takes as its input a \texttt{json} file of various parameters (e.g. diameters, number of elements, which detectors to implement, etc.) and exports a \texttt{GDML} file which is then read in by the Geant4 simulation. This file contains a full description of the physical geometry of the detectors, as well as their material properties (density, reflectivity, scintillation yield, etc.). 
Because of this workflow, it becomes trivial to implement scans over various parameters to perform systematic studies. 

\begin{figure}[H]
   \centering
   \includegraphics[width=0.35\textwidth]{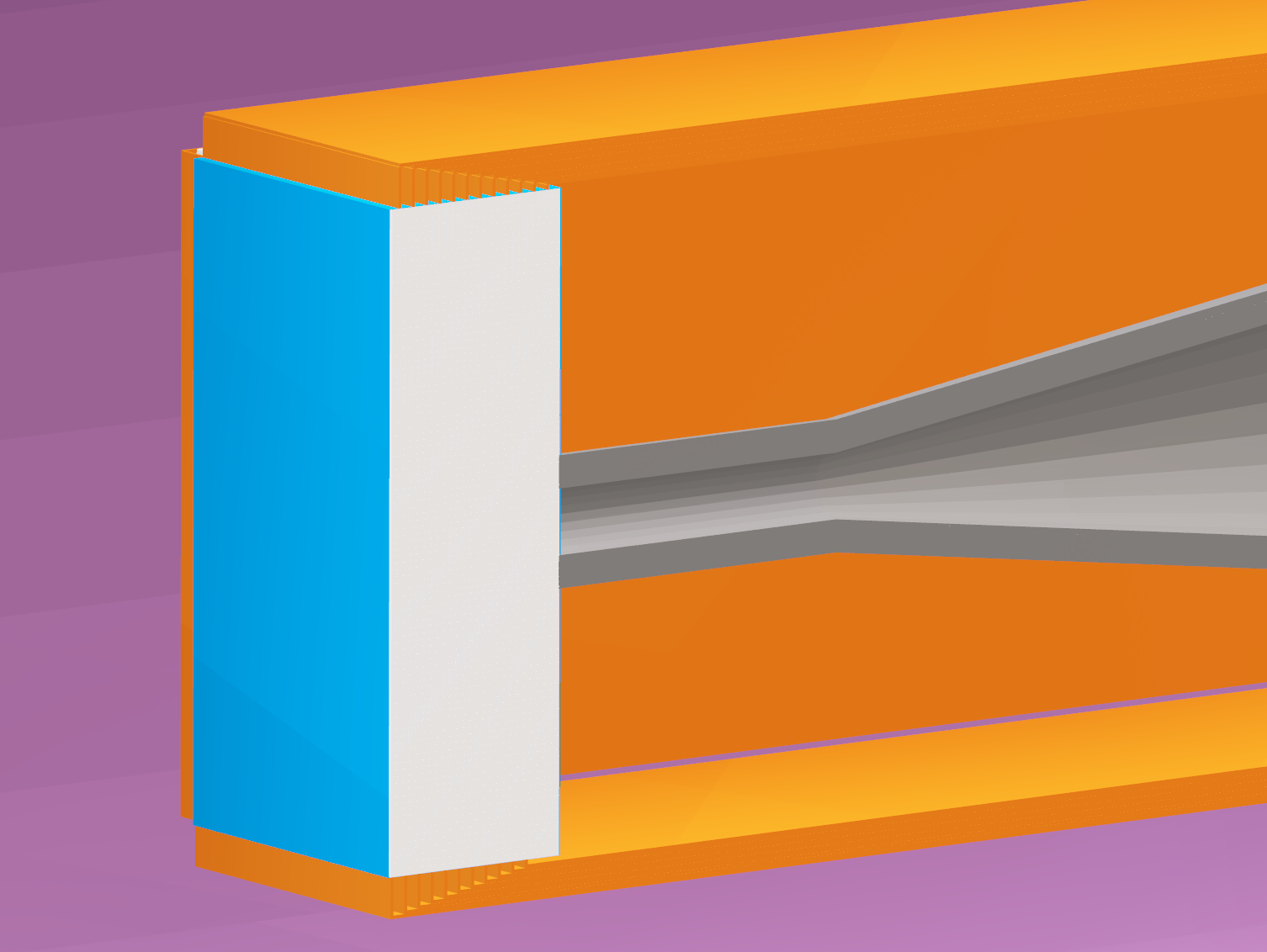}
   \includegraphics[width=0.48\textwidth]{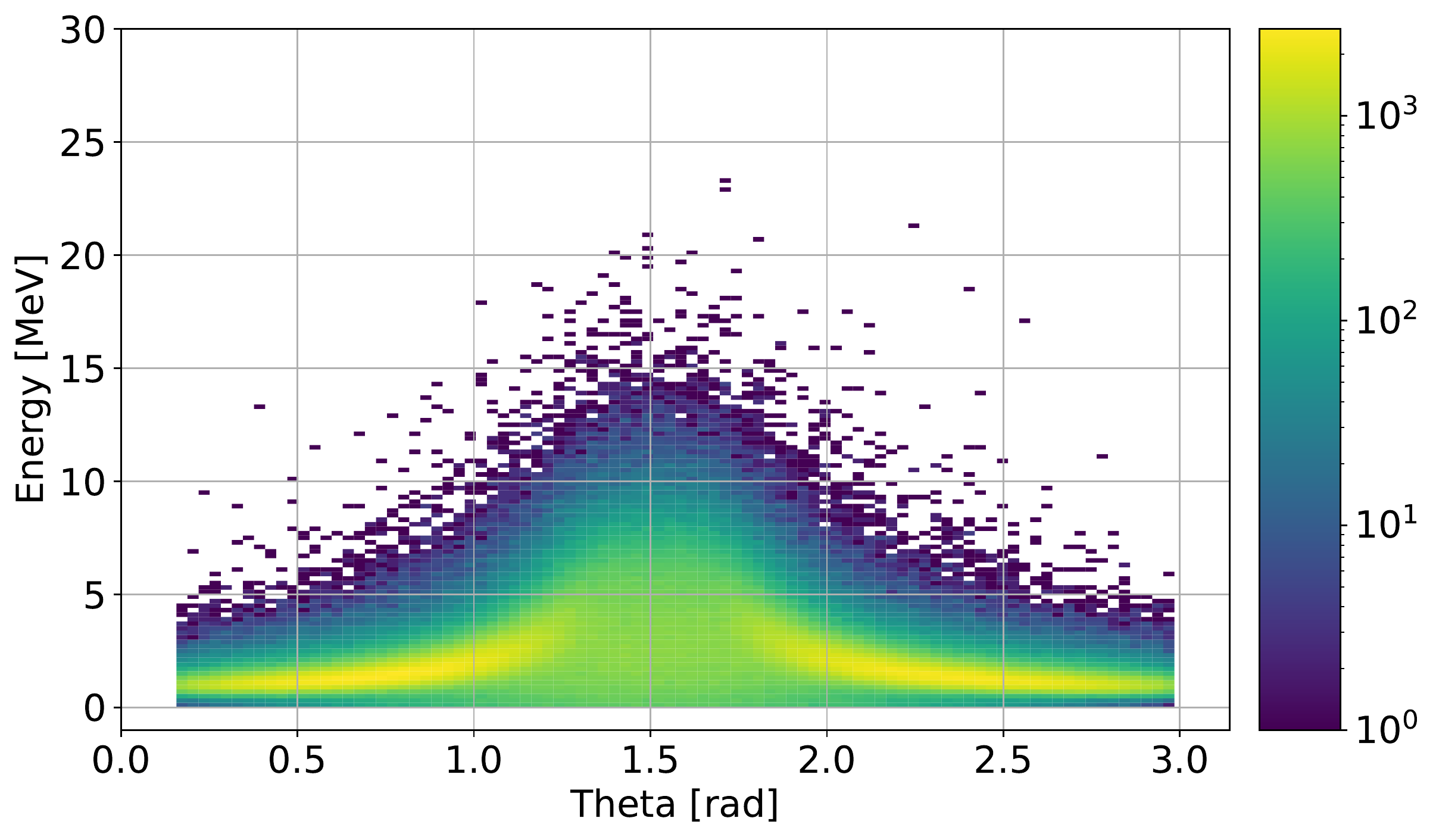}
   \caption{Left: An image of the simulated ATAR, showing the active planes (white) and readout strips (orange). Right: Total energy deposited in the ATAR by the decay positron as a function of the angle $\theta$ at which it enters the calorimeter. }
  \label{fig:atar_sim}
\end{figure}

\subsubsection{ATAR Simulation}
Simulations of the active target are done with both GEANT4, which allows us to model pions and their decay products through the full chain of detectors, and 
TRIM\cite{ziegler1985stopping}, 
which allows us to track particle energy deposition precisely in a simulated detector readout. 
Figure \ref{fig:atar_sim} (left) shows the ATAR concept  and (right) a simulation of energy deposit by exiting positrons as a function of polar angle of emission.

With an anticipated tail fraction $f= 0.5\%$ of \pie\ events, $\mu-e$ decays must be suppressed to a level that allows measurement of $\Delta f/f$  with accuracy 1--2\%. This can be accomplished by using ATAR information to identify stopped pions (and reject incoming muons) by energy loss, and range-energy relations, using a narrow time window (e.g. 3--35\,ns) following the pion stop, rejecting events with observation of the \tmu\ decay muon, and suppressing events with pion decay-in-flight (\pdif), and muon decay-in-flight (\mdif) following pion decay-at-rest (\pdar) by tracking and energy loss measurements. Experience from other experiments and PIONEER simulations indicate that $\pi\to\mu\to e$ background in the tail region from the dominant pion decay at rest will be suppressed to a negligible level, radiative decays $\pi\to\mu \nu \gamma$ followed by $\mu\to e$ decays which may leave $<4.1$ MeV in the target will be suppressed by observation of the gamma and  detection of the muon pulse and residual energy, and \mdif\  can be suppressed to about $10\%$ of the tail fraction $f$.

Another potentially significant low energy background in the tail region comes from \pdif\  in the ATAR followed by muon  decay at rest. This component dominated the background suppressed spectrum in PIENU. Simulations indicate that 0.1\% of pions decay in flight in the ATAR before stopping and initial studies indicate that the \pdif\  events can be suppressed by a factor of 5000 using ATAR tracking information which recognizes kinks in the topology and measures dE/dx along the track. Along with suppression of muon decays by selecting a narrow  time window e.g.  3--35\,ns after the pion stop, the estimated \pdif\ contribution to the uncertainty in the tail correction $f$ is negligible. 

There is an ongoing effort to apply machine learning tools to boost the sensitivity to $\pi^{+}\rightarrow e^{+}\nu$ events in the tail region and suppress background. It has already been demonstrated with the in-ATAR \pdif\  events that gradient boosting decision trees (BDT) are able to  outperform the manual cut-based methods. Although the BDT model shows excellent classification performance, there are further potential gains from deep neural networks, especially Convolutional Neural Networks (CNN), which have shown extraordinary performance in image processing and classification applications.

\subsubsection{CALO Simulation}
\begin{figure}[H]
  \centering
  \subfloat[]{\includegraphics[width=0.22\textwidth]{
  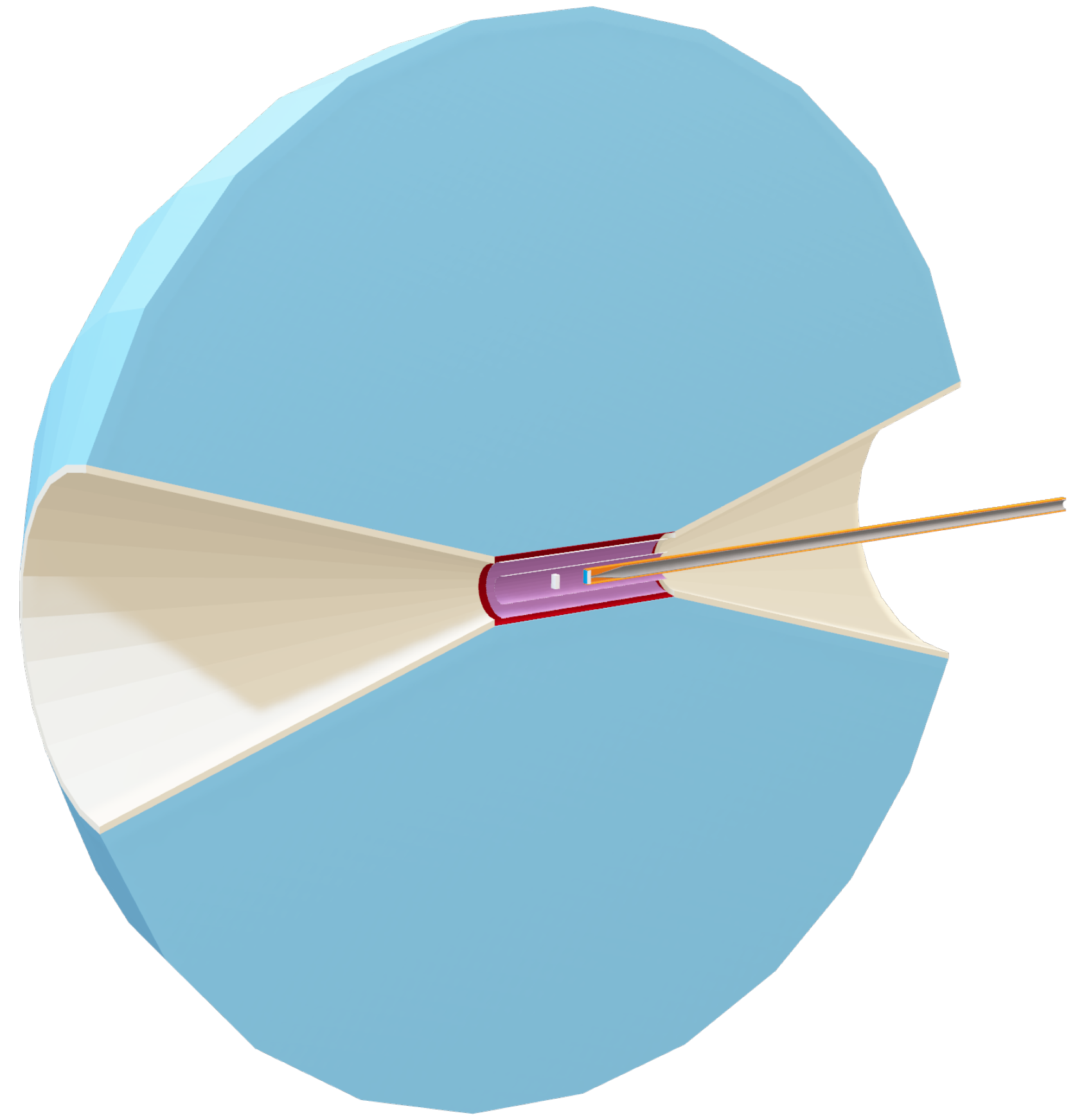}\label{fig:geometry_MC}}
  \hfill
%   \subfloat[]{\includegraphics[width=0.7\textwidth]{Figures/PSF_new.pdf}\label{fig:PSF}}
  \subfloat[]{\includegraphics[width=0.7\textwidth]{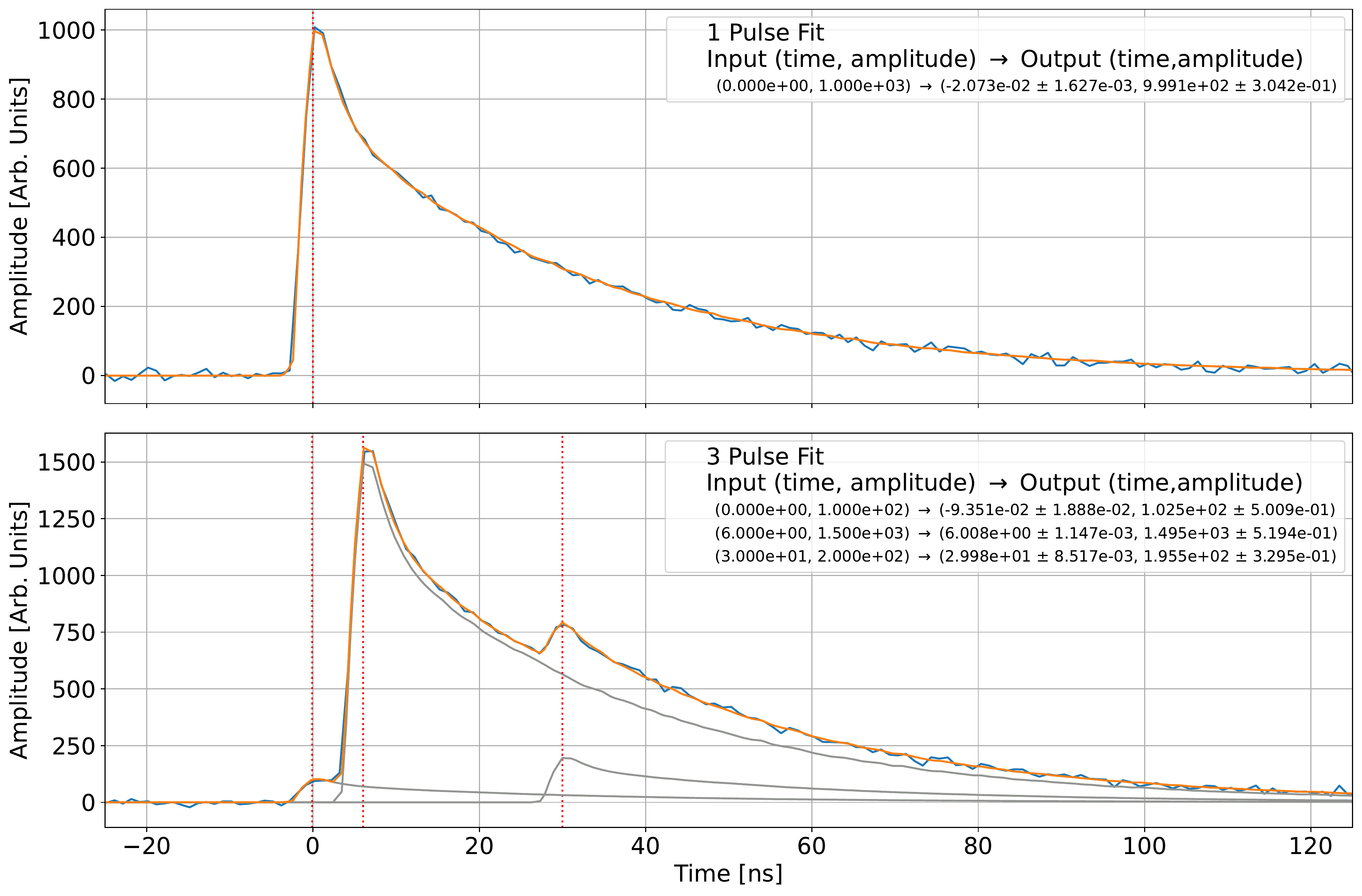}\label{fig:PSF}}
  \caption{(a) Rendering of the simplified geometry used in the first Monte-Carlo simulation of the PIONEER LXe calorimeter (b) Example of simulated pulse shapes of a single $\pi\rightarrow \mu \rightarrow e$ event (top) and 3 events happening closely in time (bottom) and recorded in the LXe calorimeter. Pulse shape fitting based on a template allows accurate identification and energy reconstruction of multiple pulses down to $\pm$\unit[5]{ns} separation.} % add some caveat
\end{figure}

Simulation studies of the LXe calorimeter response have been performed using the GEANT4 package \cite{GEANT4:2002zbu} with optical photon tracking. A simple geometry (see Fig.~\ref{fig:geometry_MC}) that is representative of the current design was implemented. $\pi\rightarrow \mu\rightarrow e$ events were generated in the target at the center of the LXe sphere and optical photons originating from LXe scintillation induced by the showers generated by the positrons were tracked until the outside sensitive surface of the LXe sphere. 
%The  simulated 
%pulses obtained are similar to the ones reported by  MEG.
A pulse fitting algorithm was employed to evaluate the possibility of separating events that overlap in time i.e.~pulse  pileup. Pulse separation down to \unit[5]{ns} was achieved (see illustration in Fig.~\ref{fig:PSF}) across a wide range of amplitudes. This gives a first indication (without digitization) of the performance of the detector and subsequent data analysis with respect to dealing with  pileup.

Further studies are anticipated to introduce and optimize optical qualities of the surfaces, optimize photo-sensing detector coverage,
and improve the simulation of the originating scintillation photons. Much  of the input for these simulations will be  provided by test measurements. 
Position resolution will  also augment  pile-up handling capabilities. The position resolution capabilities of the detector and its importance for achieving the targeted rate will be modeled  and studies  are envisaged in the apparatus which will also be used to test the photo-sensors in LXe.  
Depending on the outcome of the pileup studies, segmentation of the LXe volume may be considered.

One important aspect of the calorimeter design that is being informed by the simulations is understanding how energy from a decay positron is `lost' before it reaches the calorimeter. While energy lost in the ATAR is measured,  losses occurring in inactive material  e.g. the ATAR cabling, and calorimeter entrance windows, depend on the angle of emission.    Simulations are being used to study these effects on the positron energy  resolution.

Another important study involves modeling of photonuclear interactions. As decay positrons interact with atomic nuclei in the simulation, they will occasionally cause a nucleus to enter into an excited state with single or multiple neutron emission. When these neutrons escape the calorimeter without depositing their energy the observed energy is shifted down by multiples of the neutron binding energy (see Fig.~\ref{fig:Signal-Tail}). 
%Photonuclear processes were studied in the PIENU experiment to determine the impact on the NaI(Tl) calorimeter lineshape.  The results were used as input to the calorimeter simulations. 
Modeling and prototype studies will be pursued for evaluating these effects in LXe.

%\subsection{Beam}
%\subsubsection{Beam requirements and target setup}
%\input{PioneerExperiment/Beam/beam1}
%\subsubsection{Beam line at $\pi$E5}
%\input{PioneerExperiment/Beam/beam2}
%\subsubsection{Beam optimization and studies} \label{sec:beam_studies}
%\input{PioneerExperiment/Beam/beam3}

%\subsection{Beam Instrumentation }

%\subsection{Active target (ATAR)}
%\input{PioneerExperiment/ATAR/ATAR}
%\input{PioneerExperiment/ATAR/ATAR_appendix_short}

%\subsection{Cylindrical Tracker}
%\input{PioneerExperiment/ATAR/positron_detectors}
%\subsection{LXe Calorimeter} 
%\subsubsection {Overview}
%\input{PioneerExperiment/Calorimeter/caloverview}
%\subsubsection {Photosensors}\label{sec: LXe_photosensors}
%\input{PioneerExperiment/Calorimeter/Photosensors}
%\subsubsection{Cryogenics and Purification}\label{sec:LXe_Cryogenics-purification}
%\input{PioneerExperiment/Calorimeter/cryogenics-purification}
%\subsubsection{Calorimeter Mechanical Engineering}\label{sec:Calorimeter_Mech_Eng}
%\input{PioneerExperiment/Calorimeter/Calorimeter_Mech_Eng}
%\subsubsection{Other Calorimeter Options}
%\input{PioneerExperiment/Calorimeter/Othercal}

%\subsection{Trigger and Backend electronics}
%\input{PioneerExperiment/Trigger/trigger}
%\input{PioneerExperiment/Trigger/electronics}
%\subsection{DAQ}
%\label{subsec:daq}
%\input{DAQ/daq.tex}

% \section{Simulations and Analysis}
% \subsection{Calorimeter Simulations}
%\input{Simulations and Analysis/Geant4simulations}
%\input{PioneerExperiment/BeamSim}

\section{Sensitivity}
The sensitivity of PIONEER has been evaluated using performance assumptions that are chosen to be compatible with the experimental conceptual design described above.
\subsection {$\pi\to e \nu$}
The first phase of \nexp ~ ($R_{e/\mu}$) will employ a beam with a pion stopping rate in the ATAR of approximately $3\times 10^5$/s
  at momentum of $55-70$~MeV/$c$ with $\frac{\Delta p}{p}\leq 2\%$
in  a spot size  $\leq2$\,cm diameter. Muon and positron contaminations will be reduced to $<10\%$ with the use of a separator as discussed above. 
%These requirements are compatible with the beams available at the $\pi$E5 (and $\pi$E1) beamline. 
To estimate
%For the basis of estimating 
the running time required to reach the proposed sensitivity, we assume that the beam will be available during 5 months per year. We also assume an overall event acceptance efficiency of 30\%, which is based on the fiducial volume, the timing window cuts, and reconstruction factors.  We assume a data-taking (operations) efficiency of 50\% based on the product of PSI beam delivery and experimental data-collection uptime, along with an allocation for non-production systematic uncertainty tests.  These factors result in $2\times 10^8$ $\pi^+\to e^+ \nu$ events for a 3-year run satisfying the statistics goal. 

\begin{table}[htbp]
\centering
\begin{tabular}{lrr}
\toprule
& PIENU 2015 & PIONEER Estimate\\
Error Source & \% & \% \\
\hline
Statistics & 0.19 & 0.007 \\
Tail Correction & 0.12 & <0.01 \\
$t_0$ Correction & 0.05 & <0.01 \\
Muon DIF & 0.05 & 0.005 \\
Parameter Fitting & 0.05 & <0.01 \\
Selection Cuts & 0.04 & <0.01 \\
Acceptance Correction & 0.03 & 0.003 \\
{\bf Total Uncertainty} & {\bf 0.24} & {\bf $\leq$ 0.01} \\
\bottomrule
\end{tabular}
\caption{$Br(\pi\to e\nu)$ precision for PIENU 2015\cite{PiENu:2015seu} (left) and estimated precision for PIONEER (right).}
\label{pienu_precision}
\end{table}

%\begin{figure}[htbp]
%\centering
%\includegraphics[scale=0.5]{Figures/pienu_sensitivity.pdf}
%\caption{ $Br(\pi\to e\nu)$ precision for PIENU 2015 (left) and estimated precision for PIONEER (right).}
%\label{pienu_precision}
%\end{figure}

Systematic uncertainties for PIONEER have been estimated based on the experience of PIENU~\cite{PiENu:2015seu} and are shown in Table~\ref{pienu_precision}. The main systematic uncertainty for  PIENU was the uncertainty in the  tail correction for $\pi\to e\nu$ events below 52\,MeV. In \nexp ~the tail will be reduced from 3\% to $0.5\%$ due to the increased thickness of the calorimeter (25$X_0$ compared to $\leq$19$X_0$) and the more uniform acceptance due to the larger solid angle. The ATAR will allow suppression of \pdif\  and \mdif\  backgrounds enabling more precise measurement of the tail. Uncertainties in the other small corrections, e.g. the pion stop time ($t_0$) Correction, Selection Cuts, and Acceptance Correction, are estimated to be reduced due to the improvements such as in the calorimeter and ATAR timing resolutions. An additional uncertainty arises from the pion lifetime, presently known to 0.02\% precision~\cite{ParticleDataGroup:2020ssz}; the \nexp~ group intends to make additional measurements to reduce this uncertainty to $<0.01\%$.
\subsection {Exotics}
\subsubsection{Massive neutrino searches $\pi^+ \to \ell ^+ \nu _H$}

Searches for peaks in the positron energy spectum due to  $\pi^+ \to \ell ^+ \nu _H$ decays were performed in the PIENU experiment \cite{PIENU:2017wbj,PIENU:2019usb} sensitive to masses $65<m_H<135$ MeV  but no significant signal above statistical uncertainty was found. The measurement of $R_{e/\mu}$\cite{PiENu:2015seu} provides limits for $m_H<65$ MeV.
To estimate the expected sensitivities for PIONEER with $100\times$ the statistics (Phase I), reduced backgrounds, and improved detectors, toy MC simulations were performed.

%\subsubsection{$\pi^+ \to e^+ \nu_H$ search}\label{pie}

\begin{figure}[htbp]
\centering
\includegraphics[scale = 0.3]{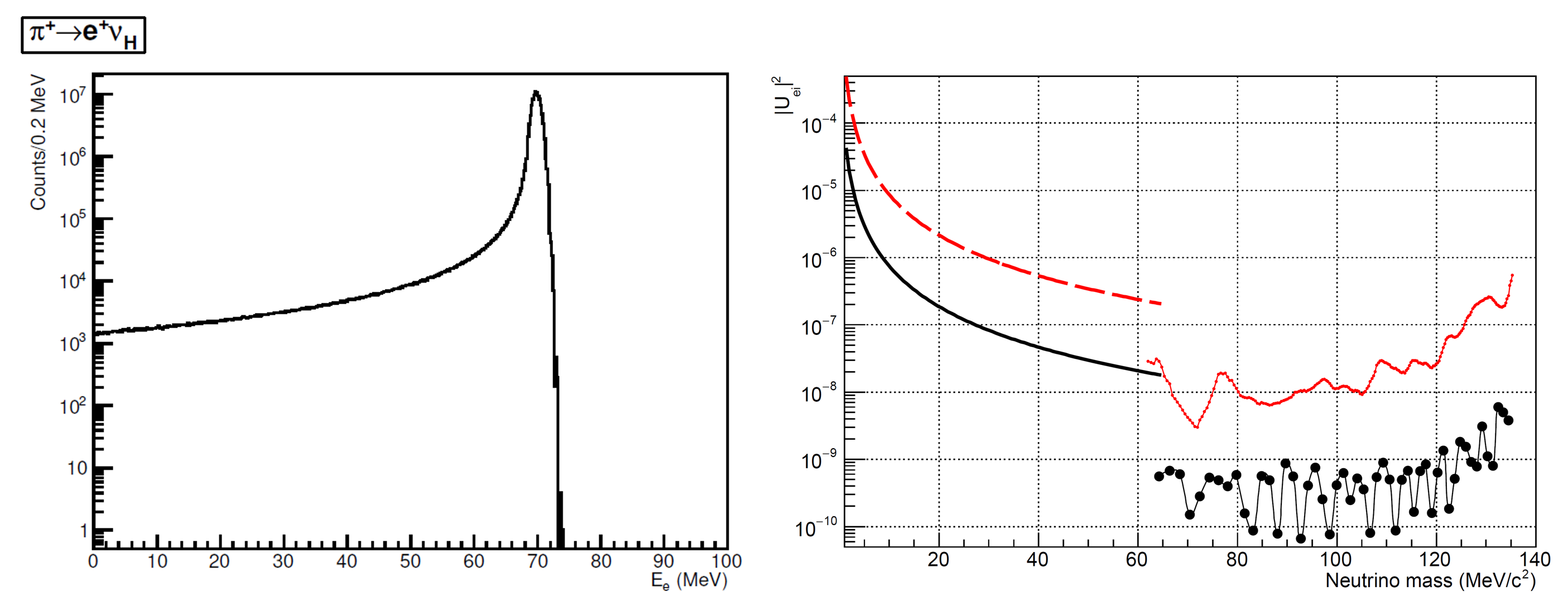}
\caption{Left: simulated $\pi^+ \to e^+ \nu_e$ energy spectrum. Right: Limits on  $|U_{e i}|^2$ (90\% C.L.) from PIENU (red)\cite{PiENu:2015seu,PIENU:2017wbj}) and expected sensitivity from  PIONEER (black). 
The  lower region limits ($m_H$<65 MeV) come from the branching ratio measurement and the  higher ($m_H$>65 MeV) region from the peak search.}
\label{pienuH}
\end{figure}

For $\pi^+ \to e ^+ \nu _H$ decays, the peak search sensitivity  was limited by residual $\pi^+ \to \mu^+ \to e^+$ background from pion and muon decay-in-flight ($\pi$DIF and $\mu$DIF). 
The low energy calorimeter response tail and statistics of the $\pi^+ \to e^+ \nu_e$ decay also limits the sensitivity. 
Using an active target and a larger electromagnetic calorimeter, the background $\pi^+ \to \mu^+ \to e^+$ will  be significantly suppressed compared to PIENU, and a significantly smaller low energy $\pi^+ \to e^+ \nu_e$ tail is anticipated. 
Figure \ref{pienuH} shows the result of a toy MC study for the expected sensitivity (90\% confidence level (C.L.) upper limits) in PIONEER, assuming $1 \times 10^8$ $\pi^+ \to e^+ \nu_e$ events, 1\% tail fraction below 52\,MeV, no $\pi^+ \to \mu^+ \to e^+$ events, improved energy resolution for decay positrons with respect to the PIENU calorimeter \cite{PiENu:2015pkq}, and negligible acceptance corrections due to the larger detector acceptance. 
Compared with PIENU (red curves in the right panel in Fig.~\ref{pienuH}), the expected sensitivity in PIONEER would be improved by one order of magnitude. 

%\subsubsection{$\pi^+ \to \mu^+ \nu_H$ search}

\begin{figure}[htbp]
\centering
\includegraphics[scale=0.3]{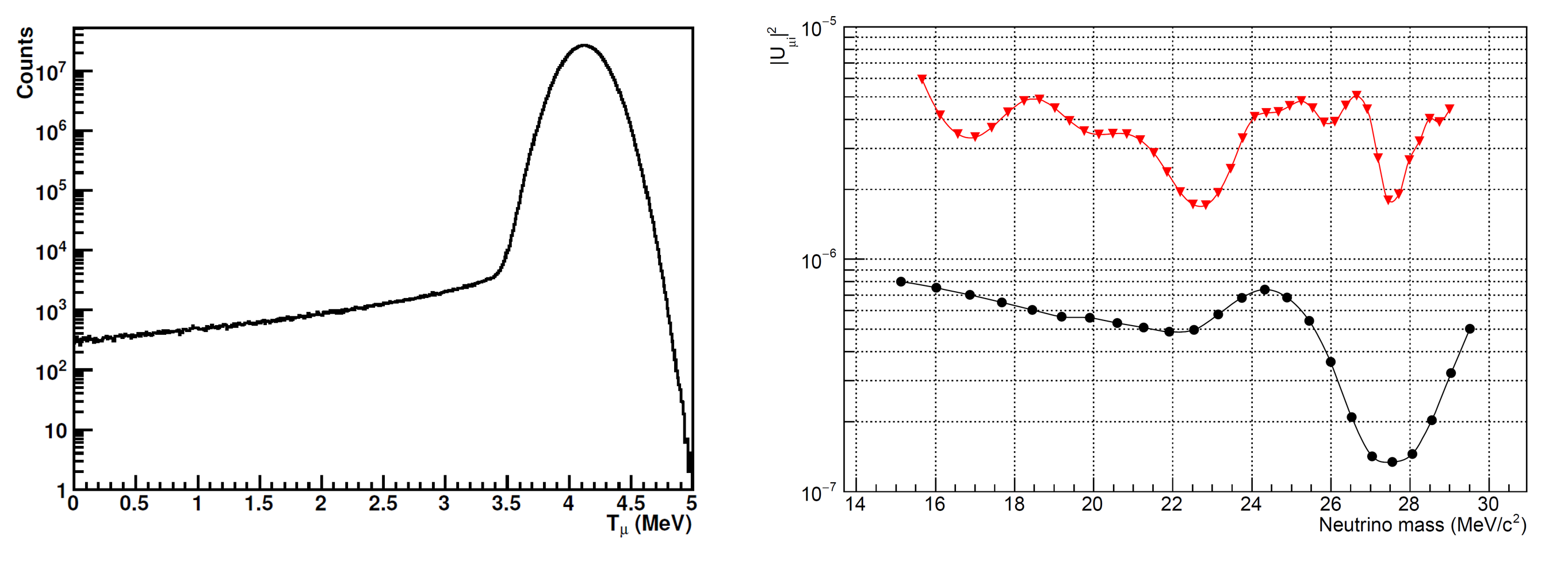}
\caption{Left: simulated $\pi^+ \to \mu^+ \nu_\mu$ kinetic energy spectrum. Right: The PIENU result (red triangles \cite{PIENU:2019usb}) and expected sensitivity with PIONEER (black circles) for the mixing matrix $|U_{\mu i}|^2$ (90\% C.L. limit).}
\label{Result}
\end{figure}

 The sensitivity for $\pi^+ \to \mu^+ \nu_H$ decay will also be improved by the larger PIONEER statistics . 
The dominant background is mainly due to the radiative pion decay $\pi^+ \to \mu^+ \nu_{\mu} \gamma$ with branching fraction  $2\times10^{-4}$ \cite{PIMUNUG} (for $E_\gamma< 1$\,MeV). 
A toy MC simulation was performed with $1 \times 10^9~{\pi}^+\to\mu^+\nu_{\mu}$ decays ($\times 100$ larger statistics than PIENU) including $\pi^+ \to \mu^+ \nu_{\mu} \gamma$ background considering the proper branching fraction, and assuming the same detector resolution as in PIENU. 
Figure \ref{Result} shows the results of the simulation and the PIENU experiment \cite{PIENU:2019usb}. 

\subsubsection{Two body muon decay $\mu^+ \to e^+ X_H$}

Massive or massless weakly interacting neutral bosons $X$ such as axions \cite{Axion1, Axion2, Axion3, Axion4} and Majorons \cite{Majoron, Majoron2, Majoron3} have been suggested to extend the SM including models with dark matter candidates, baryogenesis, and solutions to the strong $CP$ problem. 
Wilczek suggested a model \cite{Wilczek} which may lead to charged lepton flavor violation (CLFV) where the boson X can be emitted in flavor changing interactions. 

When decay products from a massive boson $X_H$ are not detected due to a long lifetime, flavor violating two-body muon decays involving a massive boson $\mu^+ \to e^+ X_H$ can be sought by searching for extra peaks in the Michel spectrum. 
This search was performed with PIENU, resulting in the limit to the branching ratio $\Gamma (\mu^+ \to e^+ X_H)/\Gamma( \mu ^+ \to e^+ \nu \bar{\nu})$ at the level of $10^{-5}$ in mass range $47.8-95.1$ MeV/$c^2$ \cite{PIENU:2020loi}. 
The  statistics will improve by two orders of magnitude with respect to PIENU. 

\begin{figure}[htbp]
\centering
\includegraphics[scale=0.5]{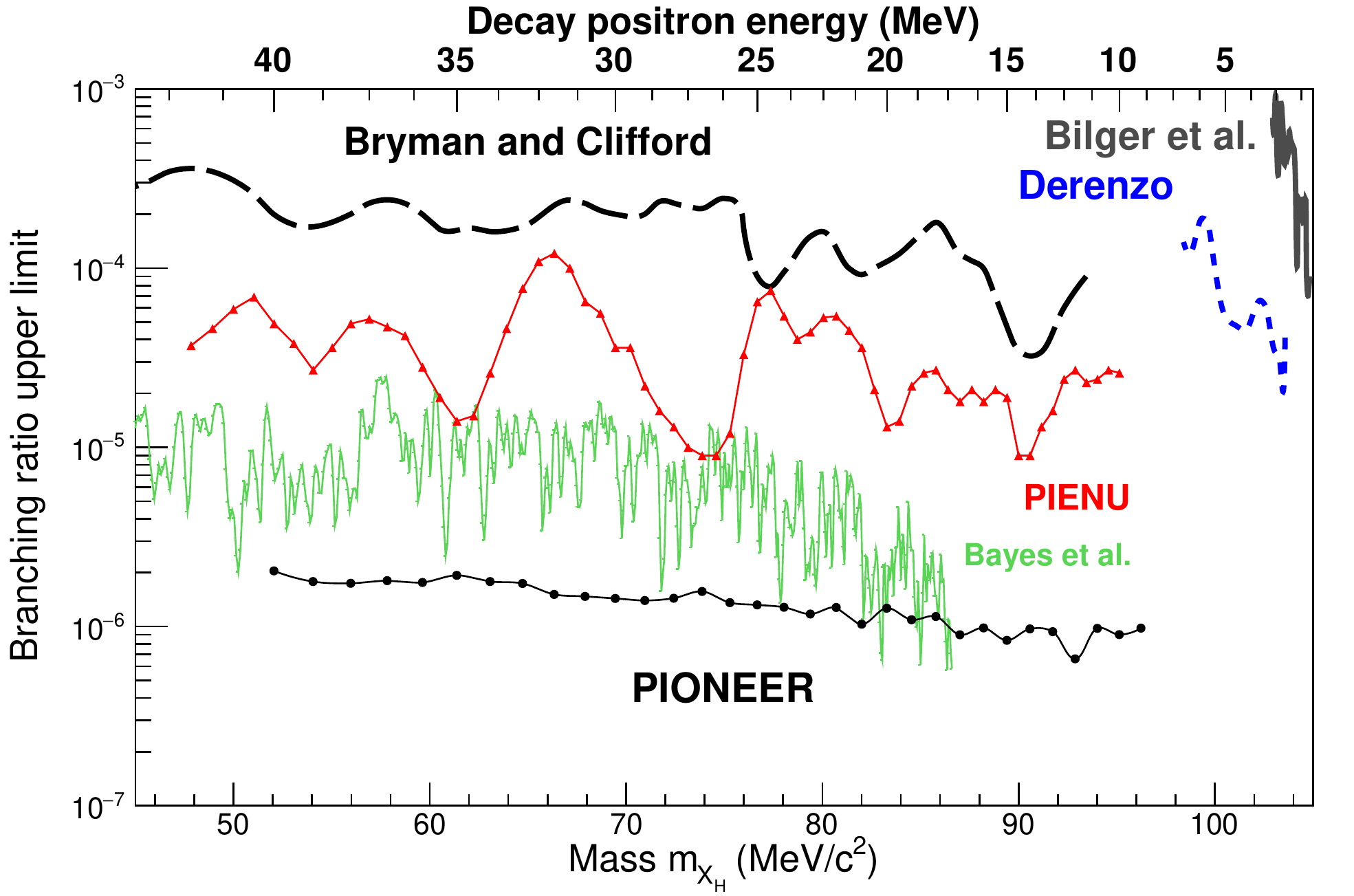}
\caption{90\% C.L. limit of the branching ratio for PIONEER (black circles) and other past experiments (see Refs. \cite{Exp1, Exp2, Exp3, Exp4} for more details).}
\label{FinalResult}
\end{figure}

To estimate the expected sensitivity, $2 \times 10^{10}$ muon decays were simulated.
%, as described in Sec. \ref{pie}. 
Figure \ref{FinalResult} shows the 90\% C.L. upper limits of the branching ratio from different experiments and the expected sensitivity for PIONEER. 

\subsubsection{Other decays}

Three and four body pion decay modes can  be analyzed with the same method as for the $\pi^+ \to \ell ^+ \nu_H$ searches. However, the signal shapes are in these cases are represented by continuous lepton energy spectra. 
The expected sensitivities will also be improved by one order of magnitude. The current limits set by the PIENU experiment are described below.

%\subsubsection{Three body pion decays $\pi^+ \to \ell ^+ \nu X$}

Three body pion decays $\pi^+ \to \ell ^+ \nu X$, where $X$ is a massive or massless weakly interacting particle, were searched for in the PIENU experiment \cite{PIENU:2021clt}. 
The decay $\pi^+ \to e^+ \nu X$ was sought; no signal beyond the statistical uncertainty was observed, and 90\% C.L. upper limits were set on the branching ratio $\Gamma (\pi^+ \to e^+ \nu X)/\Gamma (\pi^+ \to \mu^+ \nu)$ with $10^{-7}-10^{-8}$ level in the mass range of $0<m_X<120$ MeV/$c^2$. 

The $\pi^+ \to \mu^+ \nu X$ decay was also searched for.
A 90\% C.L. upper limit was derived on the branching ratio $\Gamma (\pi^+ \to \mu^+ \nu X)/\Gamma (\pi^+ \to \mu^+ \nu)$ at the $10^{-5}-10^{-6}$ level in the mass region from 0 to 33.9 MeV/$c^2$.

%\subsubsection{Rare pion decays $\pi ^+ \to \ell^+ \nu_{\ell} \nu \bar{\nu}$}

The rare pion decays $\pi ^+ \to \ell^+ \nu_{\ell} \nu \bar{\nu}$ are highly suppressed. Thus, the experimental search for these processes could reveal small non-SM effects such as neutrino-neutrino interactions \cite{nunu} and six-fermion interactions \cite{6f,6f2}, which might compete with the SM processes at first order. 
The rare pion decays, considering three models (SM, neutrino-neutrino interaction, and six-fermion) were also searched for in PIENU \cite{PIENU:2020las}, and a first result for $\Gamma(\pi^+ \to \mu^+ \nu_{\mu} \nu \bar{\nu})/\Gamma(\pi^+ \to \mu^+ \nu_{\mu})<8.6 \times 10^{-6}$ and an improved measurement $\Gamma(\pi^+ \to e^+ \nu_{e} \nu \bar{\nu})/\Gamma(\pi^+ \to \mu^+ \nu_{\mu})<1.6 \times 10^{-7}$ were obtained. 

\subsection {Pion Beta Decay}

For the $\pi^+ \to \pi^0 e^+ \nu$ experiment the  positive pion stop rate would have to be higher, $\geq 10^7$/s, possibly with  a larger momentum bite $\frac{\Delta p}{p}\approx 3\%$ and likely  using higher pion momentum. 
%These beam conditions are compatible with the $\pi$E5 beam line. 
This would result in $7\times 10^5$ $\pi^+ \to \pi^0 e^+ \nu$ events collected for 4 years of (5 months/yr) operation assuming similar efficiency factors as discussed for the $\pi\to e \nu$ measurement.\footnote{In the Phase I measurement of \pie~ \nexp\ will collect a sizeable sample of pion beta decay events, which will be helpful to inform the Phase~II (III) design.} This would be sufficient to achieve the required statistical precision to improve the pion beta decay branching ratio measurement precision  by a factor of 3 (Phase II). Systematic effects are expected to be reduced to the 0.06\% level ($10\times$ lower than for the previous PiBeta experiment) due to the combined improvements to the calorimetry (principally, the time and energy resolutions) and the ATAR which may facilitate the observation of the positron in $\pi^+ \to \pi^0 e^+ \nu$ decay in coincidence with the $\pi^0$ detection.

Running at higher rates may be possible leading to  a further precision improvement of 3 (Phase III) and will depend on the ability of the spectrometer to deal with higher rates of  pile-up of accidental events. In this regard, we are studying the possibility to optically segmenting the LXe volume.

\section{ Planning for Realization of PIONEER}
%\subsection {Collaboration}

The PIONEER collaboration consists of participants from  PIENU, PEN/PiBeta, and MEG/MEGII as well as  experts in rare kaon decays, low-energy stopped muon experiments, the  Muon $g-2$ experimental campaign, high energy collider physics, neutrino physics, and other areas. The collaboration is still developing and welcomes new members.

%We intend to  draft  a collaboration constitution and   institute an appropriate organizational structure. We have  good models, for example, from Muon g-2, that can be tailored to our smaller, but equally diverse and international, team.  We expect that there will be an elected Executive Board with a Chair that guides the key technical decisions as an advisory body to the elected Spokesperson(s) and leaders of the various important technical areas (Beam, ATAR, Calo, DAQ, etc.).  

%\subsection {R\&D }
The collaboration anticipates performing detector R\&D in several areas, including the following:
\begin{itemize}
\item Beam studies will be preformed in $\pi E5$ (and possibly $\pi E1$) to establish the required beam conditions. Two weeks of beam time has been approved for 2022.
    \item After initial sensor characterization and design optimization a PIONEER specific ATAR sensor prototype will be produced. The characterization includes studies on LGAD energy resolution and gain suppression mechanism. 
    \item A first ATAR demonstrator with a few planes of available sensor prototypes will be produced. An electronics board with suitable characteristics needs to be designed and produced. The prototype would be then tested in a pion/muon beamline.
    \item Identification of suitable chips for the ATAR analog amplification and digitization. 
    \item ATAR mock-ups: the support mechanics and thermal load will be studied with mock-up prototypes and silicon heaters.
    \item Cylindrical positron tracker. Designs with standard 300~$\mu m$ thick Si strips and with LGADs are being considered. We expect to construct and test  prototypes of various geometries.
    \item LXe prototype: The  objectives of this R\&D work include   determination of the properties of photo-sensors  and optical properties of materials for use in the  LXe calorimeter. We also want to  benchmark the photon transport simulations.  We are considering the development of a medium scale calorimeter prototype that would enable measurements of properties like energy resolution and photonuclear effects for validation of simulations.
    \item LXe calorimeter optical segmentation: Small prototypes will be used initially and UV compatible materials will be evaluated. Some of these studies may  be done using a LXe cryostat at McGill university containing $\sim$\unit[2]{l} of LXe developed  for SiPM tests for nEXO.  An assembly hosting SiPMs, reflective material and a retractable radioactive source will be prepared at TRIUMF and brought to McGill for measurements. 
    \item SiPMs: SiPM degradation at high rates will be studied. We will test available photosensors using small LXe prototypes in association with the McGill setup mentioned above.
    \item Crystal alternatives to LXe: Arrays of LYSO crystals with varying levels of doping will be evaluated from various manufacturers. %See the Appendix for more details.
    \item DAQ: Rate testing of FPGA-to-CPU/GPU and CPU-to-CPU communication via optical PCI-express links will be done along with performance testing of data compression algorithms for \calo\ data.
    \item Trigger prototyping: A prototype APOLLO Command Module will be used. A 4-channel prototype of the digitizer board for evaluation and communications development will be built.
\end{itemize}

%\subsection{ATAR R\&D}
%A brief summary of ATAR R\&D  follows.  
%A more detailed plan is presented in Appendix~\ref{sec:ATAR_timeline_app}. 
%\begin{itemize}
 %   \item After initial sensor characterization and design optimization a PIONEER specific prototype production should happen by the end of 2023. The characterization includes studies on LGAD energy resolution and gain suppression mechanism.
  %  \item Building of a first ATAR demonstrator (ATAR0) with a few planes of available sensor prototypes (BNL strip LGADs, 2.5\,cm with 500\,$\mu$m  pitch). An electronics board with suitable characteristics needs to be designed and produced. The prototype would be then tested in a pion/muon beamline either at TRIUMF or PSI. This  prototype may be produced by the end of 2023.
  %  \item Identification of a suitable chips for the analog amplification and digitization by 2024. The effect of a short flex between sensor and chip will be studied within 2022.
  %  \item The support mechanics and thermal load needs to be studied well with mock-up prototypes and silicon heaters. These details needs to be fully understood by 2025.
 %   \item Full production of sensors and readout ASIC, once identified, should take less than a year given the modest area of the ATAR. Therefore final production and subsequent assembly can start in 2025.
%\end{itemize}

%\subsection {Timeline Estimate}
%\label{sec:timeline_estimate}

An optimistic schedule for \nexp\, assuming that funding decisions are positive and proceed expeditiously, would allow 
an increasing amounts of prototype instrumentation on various sub systems (e.g. ATAR, DAQ) to be tested in the following years leading to the full-scale measurement program of Phase I to begin in 2029. 
%As seen in Table~\ref{tab:cost}, the preliminary cost estimate for \nexp\ Phase I is about 26M ChF.

%We present an optimistic schedule  for \nexp\ for the next several  years under the assumptions that program approval stages and external funding decisions are positive and proceed expeditiously: 

%\begin{itemize}%[leftmargin=*]
%\item 2022:  Submit experimental proposal to PSI and make R\&D requests to funding agencies (US, Japan, Europe, Canada);  initiate first lab tests of prototype devices; perform simulations and further develop the experiment design; beam properties test at PSI.
%\item 2023--24:  Beamline studies, detector prototype development and  test beam measurements; technical design report; funding decisions.
%\item 2025--27: Full-scale production of detectors, electronics, DAQ sub-systems; short physics integration runs of available subsystems.
%\item 2028:  \nexp\ engineering run and first physics production.
%\item 2029: Full-scale  physics measurement program.
%\end{itemize} 

%\subsection{Cost Estimate}
%\input{Planning/Cost}

%\section{Training,  Equity, Diversity, and Inclusion}
%\input{TrainingEDI.tex}

\section{Summary}
In this SNOWMASS white paper, the physics motivation and the conceptual design of a next-generation
rare pion decay experiment, PIONEER, are described. Built upon the excellent 4D tracking capability 
of the LGAD-based active target and a large-acceptance, deep, fast, and uniform liquid Xenon calorimeter with excellent energy resolution, the Phase I PIONEER experiment, approved at PSI,
aims at measuring the charged-pion branching ratio to electrons vs.\ muons $R_{e/\mu}$ 
at a precision of 1 part in $10^4$. This  precise measurement of $R_{e/\mu}$ 
can be compared with the Standard Model (SM) prediction at similar precision to probe 
non-SM explanations of several experimental anomalies pointing towards the potential violation of 
lepton flavor universality through sensitivity to quantum effects of new particles up to the PeV mass scale. 
The later phases of the PIONEER experiment aim at improving the experimental precision of the branching ratio of 
 pion beta decay, $\pi^+\to \pi^0 e^+ \nu (\gamma)$, to test CKM unitarity and to 
extract $|V_{ud}|$ at the 0.02\% level.
%The later phases of the PIONEER experiment aim at improving the experimental precision of branching ratio 
%of the pion beta decay (BRBP), $\pi^+\to \pi^0 e^+ \nu (\gamma)$, currently at 
%$\left(1.036\pm0.006\right)\times10^{-8}$, by a factor of three (Phase II) and 
%an order of magnitude (Phase III).  Such precise measurements of BRBP will allow for tests of CKM unitarity in 
%light of the Carbibo Angle Anomaly and the theoretically cleanest extraction of $|V_{ud}|$ at 
%the 0.02\% level, comparable to the currently most precise deduction from superallowed beta decays.

\begin{acknowledgements}
This document was prepared by the PIONEER collaboration in coordination with scientific staff members at PSI. The members of the collaboration are supported by the U.S. Department of Energy, Office of Science, Offices of High Energy Physics and Nuclear Physics; the U.S. National Science Foundation; Natural Sciences and Engineering Research Council (Canada); TRIUMF; the Swiss National Science Foundation; and JSPS KAKENHI (Japan).
%We gratefully acknowledge .... 

\end{acknowledgements}

%\clearpage
%\appendix
%\appendixpage
%\addappheadtotoc
%\input{Appendices/pienupenReview_appendix}
%\input{Appendices/pibetaReview}
%\input{Appendices/LYSO-alt}
%\input{PioneerExperiment/ATAR/ATAR_appendix_short}

%\input{Appendices/Authors}
%\bibliographystyle{plain} 
\bibliography{References}

\end{document}